\documentclass[manuscript]{aastex}
\usepackage{latexsym,amssymb,amsmath,amsbsy,amsopn,amstext,xcolor,multicol}
\usepackage{graphicx,wrapfig,fancybox}
\usepackage{subfigure}
\usepackage{booktabs} 
\usepackage{verbatim}
\usepackage{mathrsfs}
\usepackage{overpic,psfrag}
\usepackage{natbib}
\usepackage{multirow,tabularx}
\usepackage{epstopdf}
\usepackage{pgf}
\usepackage{rotating}

\begin{document}
\title{The star cluster mass--galactocentric radius relation:
  Implications for cluster formation}

\author{Weijia Sun}
\affil{School of Physics, Peking University, Yi He Yuan Lu 5, Hai Dian
  District, Beijing 100871, China}

\author{Richard de Grijs} 
\affil{Kavli Institute for Astronomy \& Astrophysics and Department of
  Astronomy, Peking University, Yi He Yuan Lu 5, Hai Dian District,
  Beijing 100871, China\\ and\\ International Space Science
  Institute--Beijing, 1 Nanertiao, Zhongguancun, Hai Dian District,
  Beijing 100190, China}

\author{Zhou Fan}
\affil{Key Laboratory of Optical Astronomy, National Astronomical
  Observatories, Chinese Academy of Sciences, 20A Datun Road, Chaoyang
  District, Beijing 100012, China}

\and

\author{Ewan Cameron} 
\affil{SEEG, Department of Zoology, University of Oxford, The
  Tinbergen Building, South Parks Road, Oxford OX1 3PS, UK}

\begin{abstract}
Whether or not the initial star cluster mass function is established
through a universal, galactocentric-distance-independent stochastic
process, on the scales of individual galaxies, remains an unsolved
problem. This debate has recently gained new impetus through the
publication of a study that concluded that the maximum cluster mass in
a given population is not solely determined by size-of-sample
effects. Here, we revisit the evidence in favor and against stochastic
cluster formation by examining the young ($\lesssim$ a few $\times 10^8$
yr-old) star cluster mass--galactocentric radius relation in M33, M51,
M83, and the Large Magellanic Cloud. To eliminate size-of-sample
effects, we first adopt radial bin sizes containing constant numbers
of clusters, which we use to quantify the radial distribution of the
first- to fifth-ranked most massive clusters using ordinary
least-squares fitting. We supplement this analysis with an application
of quantile regression, a binless approach to rank-based regression
taking an absolute-value-distance penalty. Both methods yield, within
the $1\sigma$ to $3\sigma$ uncertainties, near-zero slopes in the
diagnostic plane, largely irrespective of the maximum age or minimum
mass imposed on our sample selection, or of the radial bin size
adopted. We conclude that, at least in our four well-studied sample
galaxies, star cluster formation does not necessarily require an
environment-dependent cluster formation scenario, which thus supports
the notion of stochastic star cluster formation as the dominant star
cluster-formation process within a given galaxy.
\end{abstract}

\section{Introduction}

Although the initial cluster mass function (ICMF) appears to be
universal \citep{deGrijs2003,SPZ2010,Fall2012}, the formation
conditions of the highest-mass clusters are still subject to
significant debate. Proponents of one school of thought suggest that
the formation of the most massive clusters in a given volume or during
a specific time period is independent of environment
\citep{Gieles2006,Gieles2009} and, consequently, that the
``size-of-sample'' effect determines the masses of the most massive
clusters in a given cluster population (e.g., Hunter et
al. 2003). Alternatively, the formation of the most massive clusters
may require special physical conditions, such as high ambient pressure
or enhanced gas densities, which might indeed be
environment-dependent. In the context of the ICMF, its possible
environmental dependence within a given galaxy could be demonstrated
through careful analysis of its relation to galactocentric distance,
i.e., by assessment of the ``star cluster mass--galactocentric radius
relation.'' \cite{Larsen2009} modeled the ICMFs in actively
star-forming spiral and starburst galaxies, adopting a single
Schechter function, and found that the ICMF's chracteristic cut-off
mass at the high-mass end is higher for starburst galaxies than for
more quiescent spiral galaxies resembling the Milky Way. This suggests
that this global difference may be owing to the high-pressure
environments in starburst galaxies.

Pflamm-Altenburg et al. (2013; henceforth PA13) explored the apparent
cluster mass--galactocentric radius relationship of young star
clusters in Messier 33 (M33, the Triangulum galaxy) based on the data
set of \cite{Sharma2011}. They claimed to have ruled out the
size-of-sample effect as a mechanism driving massive cluster formation
and concluded that very massive star clusters may indeed require
special physical conditions to form, and thus that the formation of
young star clusters is not a stochastic process. PA13's claims appear
strong, in the sense that they purport to expose deficiencies
apparently perpetrated by the authors of a number of previous analyses
(see also Cameron 2013). However, the scientific strategy employed in
these previous studies appears entirely reasonable: namely, their
authors compare observational data against a null distribution model
and, in the event of observing a reasonable agreement, they decided
not to reject the null. Indeed, the quote from past research
highlighted by PA13 shows that the earlier researchers were careful to
avoid the common logical fallacy of treating non-rejection as support
for the null: ``Our conclusion [in support of the random drawing
  hypothesis] remains provisional'' (the text within square brackets
is ours). Unfortunately, however, the data set used by PA13 is not
suitable for the purposes of their analysis (for evidence, see Section
\ref{m33data.sec}). The validity of their conclusions must thus be
revisited.

In this paper, we use observations of young clusters in both Messier
51 (M51, the Whirlpool galaxy) and Messier 83 (M83) to independently
test PA13's conclusions. We do not only use the ordinary least-squares
(OLS) method employed by these latter authors, but we also introduce
``quantile regression'' (QR) as a more suitable and highly robust
statistical approach to support our conclusions. We focus
predominantly on the results for M51, since they are the strongest. In
Section \ref{Data}, we explain why we discard all currently available
cluster data sets pertaining to $\la 10$ Myr-old clusters in M33
(which corresponds to the age range selected by PA13) and select the
cluster samples in M51 and M83 instead (but see Section \ref{LMC} for
a brief revisit of the M33 cluster population). We also apply our
analysis approach to the lower-mass cluster sample in the Large
Magellanic Cloud (LMC). In Section \ref{M51}, we apply both methods to
M51 and show that any radial dependence is rather weak, a statement
for which we present statistically robust evidence. In Section
\ref{Parameter}, we discuss the dependence of our results on our
adoption of the maximum cluster age, minimum cluster mass, and bin
size. We conclude that variation of the lower-mass limit has a
significant impact on the results, whereas varying the applicable age
range is less important. We follow up this analysis with similar
analyses applied to the M83 as well as to the LMC and M33 cluster
populations in Sections \ref{M83} and \ref{LMC}, respectively. In
Section \ref{Conclusion}, we briefly summarize the paper and restate
our main conclusions.

\section{Observational Datasets}
\label{Data}

Our analysis relies on having access to statistically significant
numbers of young clusters with reliable age and mass estimates. In
addition, we require the host galaxies to exhibit a reasonable
symmetry so that radial averaging can be performed appropriately. In
the local Universe, the most suitable star cluster samples that are
readily available for our purpose include those hosted by M33, M51,
and M83, as well as the LMC cluster population (although the latter
galaxy is not as symmetrical as its larger spiral counterparts).

\subsection{The M33 Cluster Population}
\label{m33data.sec}

PA13 based their results for M33 on the $24 \mu$m-selected data set of
\cite{Sharma2011}. Therefore, as a first step, we attempted to
retrieve this database ourselves in order to check their
results. However, we soon learned that \cite{Sharma2011}'s ``cleaned''
dataset of 648 objects has not been made publicly available, since it
consists of an inhomogeneous mixture of object types, including
genuine young clusters, but also H{\sc ii} regions and unidentified
asymptotic giant branch stars. Nevertheless, and despite these
concerns (voiced by the original authors themselves),
\cite{Sharma2011} proceeded to apply statistical analysis tools to
both their original, highly contaminated sample of 915 sources and the
cleaned data set, referring to the latter as the ``young star cluster
sample.'' Upon contacting these authors, we learned that the
individual cluster ages and masses would be insufficiently accurate
for the type of analysis we intended to embark on (S. Sharma 2014,
private communication). They recommended us not to use their data for
exploration of the cluster mass--galactocentric radius relation, given
that they were well aware of significant, persistent contamination of
even their cleaned data set by spurious, non-cluster objects. However,
these concerns were not well communicated in their peer-reviewed
article, which led PA13 to chase after what transpired to be a poorly
defined, inhomogeneous dataset.

We therefore explored alternative databases containing cluster age and
mass estimates for significant samples of M33 clusters. A subset of
the current authors published the current most up-to-date and most
carefully defined M33 cluster data set \citep{Fan2014}. The latter
sample of star clusters was mainly selected from \cite{Roman2010},
whose database is, in turn, based on observations with the MegaCam
camera mounted on the Canada--France--Hawai'i Telescope. Fan \& de
Grijs (2014) derived photometric and physical parameters for 588
clusters and cluster candidates from archival $UBVRI$ images from the
Local Group Galaxies Survey. Supplemented by 120 confirmed star
clusters from the updated, 2010 version of \cite{Sarajedini2007}, the
total number of clusters in their sample reached 708. However,
although 69\% of these clusters and cluster candidates are
characterized by ages younger than 2 Gyr, few young clusters with ages
of $\la 10^7$ yr are included in the final database. This limitation
implies that this database cannot be used either to conclusively
examine the formation modes of young star clusters following PA13. We
are not aware of any other M33 cluster database that could currently
serve this purpose.

\subsection{M51 and M83 Cluster Data}

The M51 dataset used in this paper, which contains many young star
clusters, was published by \cite{Chandar2011}. These authors made use
of multi-color F435W (``$B$''), F555W (``$V$''), F814W (``$I$'') and
F658N (``$\text{H}\alpha$'') images of M51 obtained with the {\sl
  Hubble Space Telescope}'s ({\sl HST}) Advanced Camera for Surveys,
combined with F336W (``$U$'') pointings obtained with the {\sl HST}'s
Wide-Field and Planetary Camera-2. They estimated the age and
extinction pertaining to each cluster by performing a minimum-$\chi^2$
fit, comparing the measurements in five filters ($UBVI$,
$\text{H}\alpha$) with predictions from the Bruzual \& Charlot (2003)
simple stellar population (SSP) models. They assumed solar
metallicity, $Z = 0.02$ (which is appropriate for young clusters in
M51; Moustakas et al. 2010), a Salpeter (1955) stellar initial mass
function (IMF), and a Galactic-type extinction law. They constructed
the mass function (MF) of 3812 intermediate-age, (1--4)$\times 10^8$
yr-old star clusters in a $3 \times 7$ kpc$^2$ region of M51. For
reasons of consistency and comparison with previous work, here we
adopt Chandar et al.'s (2011) $\sim 90$\% completeness limit of $\sim
6000 M_\odot$ (but see Section 5.2 for a detailed analysis).

For M83, we adopted the dataset taken from \cite{Bastian2011}. These
authors used Early Release Science data of two adjacent fields in M83,
observed with the {\sl HST}/Wide Field Camera-3. The data pertaining
to the inner field were presented by \cite{Chandar2010}. The latter
authors made use of observations in the F336W ($U$), F438W ($B$),
F555W ($V$), F657N (H$\alpha$), and F814W ($I$) filters. The
outer-field data were imaged in the same filters. \cite{Bastian2011}
estimated the age, mass, and extinction affecting each of their M83
sample clusters by comparing the observed cluster magnitudes with SSP
models. They compared the results from their application of two
methods, including those returned by the 3DEF fitting code
\citep{Bastian2005} combined with the {\sc galev} SSP models, as well
as results obtained with the fitting procedure of
\cite{Adamo2010a,Adamo2010b} and the {\sc yggdrasil} SSP models
\citep{Zackrisson2011}. \cite{Bastian2011} adopted a metallicity of
$2.5 Z_\odot$ and a Kroupa-type stellar IMF. Both sets of models also
include contributions from nebular emission, which is helpful in
distinguishing old clusters from young, highly extinguished clusters
\citep{Chandar2010,Konstantopoulos2010} and contributes to the
broad-band colors, sometimes significantly so (Anders \&
Fritze-v. Alvensleben 2003). Both methods yielded consistent ages and
masses for the clusters. In the present paper, we have adopted the
ages and masses for 940 star clusters and cluster candidates in M83
derived using the \cite{Adamo2010a} method.

\subsection{Star Clusters in the LMC}

The statistically complete LMC cluster database we adopted was
published by \cite{Baumgardt2013}. These authors used four recent
compilations of LMC star cluster parameters to derive a combined
catalog of LMC clusters. Their primary data set is that of
\cite{Glatt2010}, who used data from the Magellanic Clouds Photometric
Surveys \citep{Zaritsky2002,Zaritsky2004}, combined with isochrone
fitting, to derive ages and luminosities for 1193 populous star
clusters in a 64 deg$^2$ area of the LMC. Their second data set is the
catalog of \cite{Pietrzynski2000}, who used $BVI$ data from the
Optical Gravitational Lensing Experiment II ({\sc ogle ii})
\citep{Udalski1998}, again combined with isochrone fitting to derive
ages of approximately 600 star clusters located in the central parts
of the LMC, with ages younger than 1.2 Gyr. \cite{Baumgardt2013} also
include cluster ages from both \cite{Milone2009} and
\cite{Mackey2003}. Many of the ages in these catalogs were derived on
the basis of {\sl HST} data.

\section{Importance of a ``truncation mass''}
\label{truncation}

Numerous previous ICMF determinations have also found power-law
functions with indices close to $-2$ (e.g., Zhang \& Fall 1999; de
Grijs et al. 2003; Bik et al. 2003; McCrady \& Graham 2007; among many
others). In addition, there is mounting evidence of the reality of a
``truncation mass'' in cluster MFs which varies among galaxies (e.g.,
Gieles et al. 2006; Bastian 2008; Larsen 2009; Maschberger \& Kroupa
2009; Bastian et al. 2012; Kostantopolous et al. 2013; Adamo \&
Bastian 2015). Indeed, the functional form of the ICMF for young star
clusters is well represented by a Schechter (1976) distribution,
\begin{equation*}
\psi(M)=\frac{\textrm d N}{\textrm d M}=AM^{-\beta}\exp(-M/M_\star),
\end{equation*}	
where $M_\star$ is the truncation mass. Gieles et al. (2006) and
Gieles (2009) reported a truncation mass that is different for spiral
and starburst galaxies. For Milky Way-type spiral galaxies, $M_\star
\approx 2 \times 10^5 M_\odot$ (Gieles et al. 2006; Larsen 2009). For
interacting galaxies and luminous infrared galaxies, Bastian (2008)
obtained $M_\star \geq 10^6 M_\odot$. In the present paper, we focus
on the importance of a possible environmental influence at different
galactocentric radii {\it in the same galaxy}. In this context,
different truncation masses in different galaxy types do not rule out
stochastic cluster formation in a given galaxy.

\begin{figure}[htbp]
	\centering
	\subfigure[]{
		\includegraphics[scale=0.25]{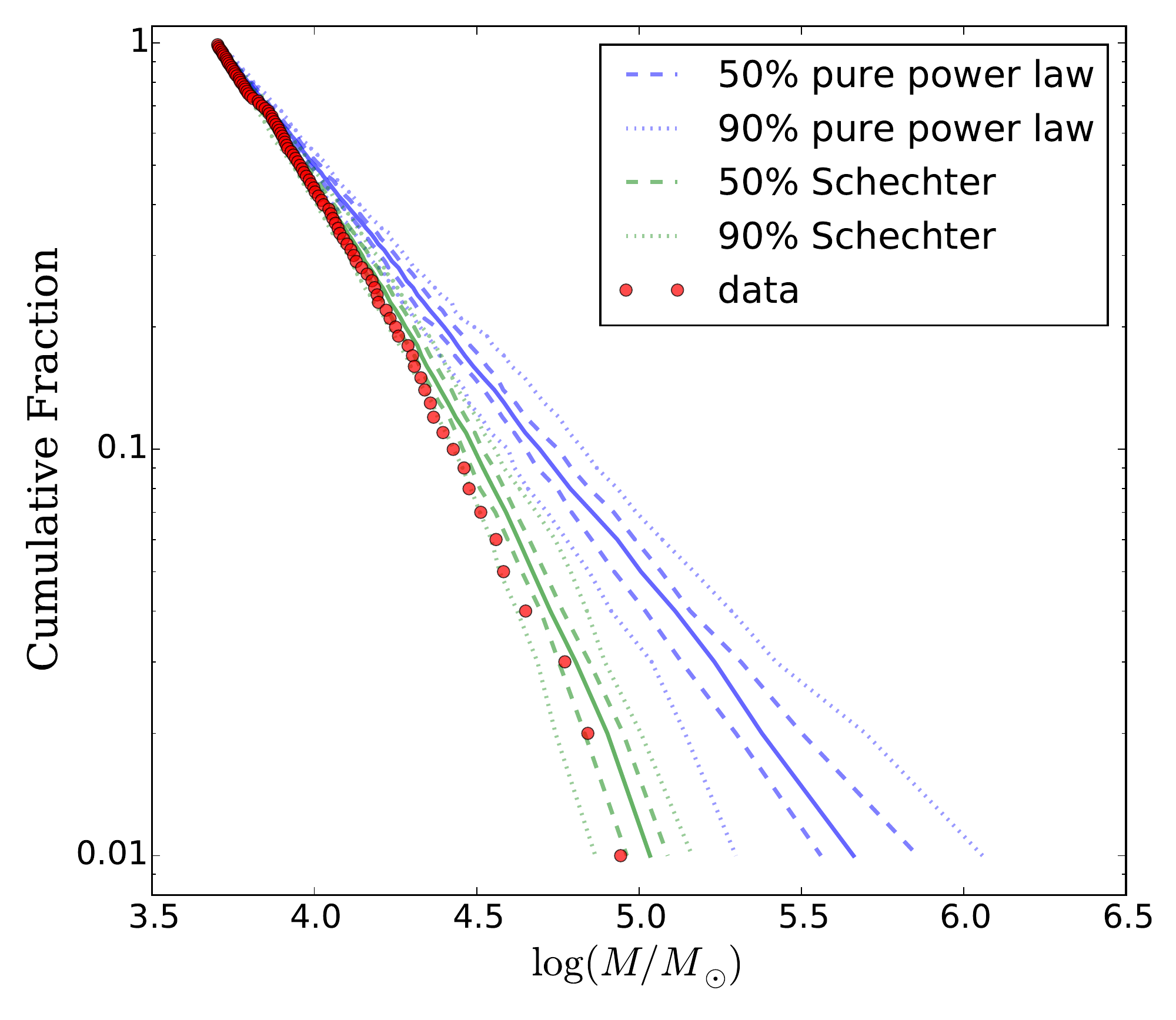}
		\label{M51_CF_7_1}
	}
	\subfigure[]{
		\includegraphics[scale=0.25]{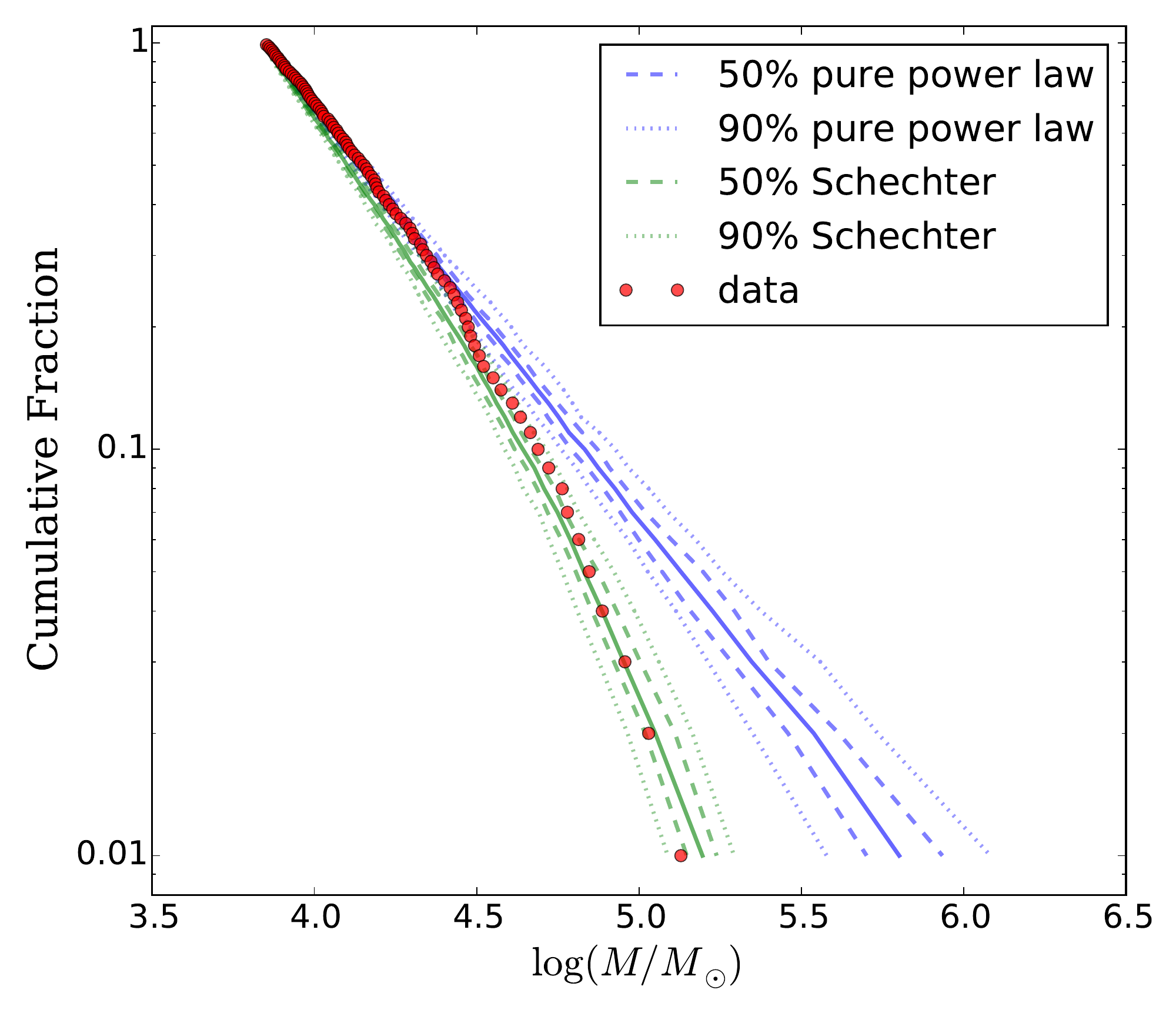}
		\label{M51_CF_8_1}
	}
	\subfigure[]{
		\includegraphics[scale=0.25]{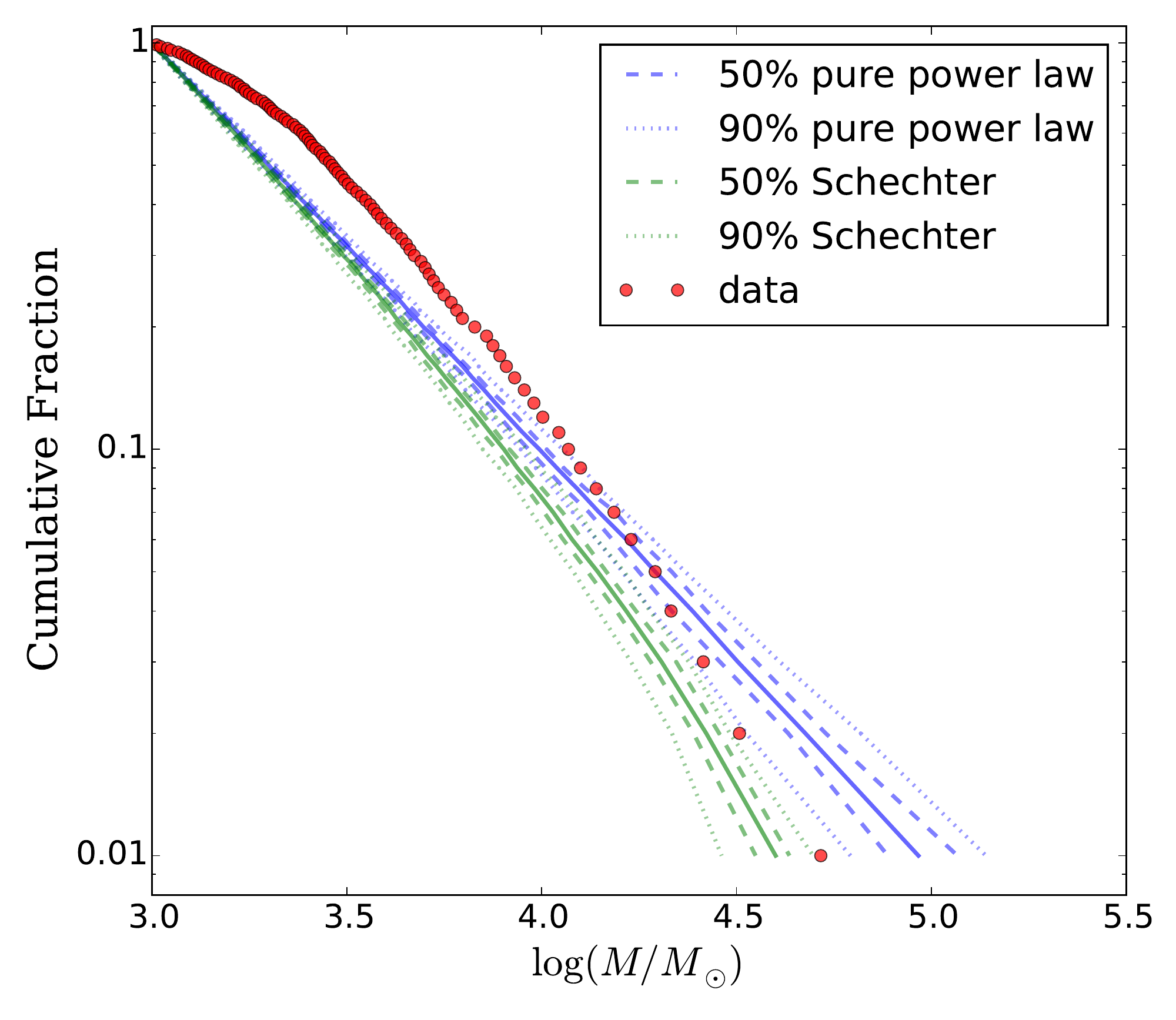}
		\label{M51_CF_7_2}
	}
	\caption{Cumulative cluster mass fractions for M51. Red dots:
          Observational data. Green line: Schechter function. Blue
          line: pure power law. (a) Age $< 10$ Myr, $M_\text{min}=5000
          M_\odot$, $\log(M_{\rm \ast}/M_\odot)=5.1$, $\beta=2$; (b) Age
          $< 100$ Myr, $M_\text{min}=7000 M_\odot$,
          $\log(M_\star/M_\odot)=5.3$, $\beta=2$; (c) Age $< 10$ Myr,
          $M_\text{min}=1000 M_\odot$, $\log(M_\star/M_\odot)=5$,
          $\beta=2$.}
	\label{CF}
\end{figure}

In previous research on the reality of truncation masses, the age of
the cluster samples used for the analyses usually extended to a few
$\times 10^8$ yr. Whether the high-mass end of the cluster MF is
affected by a truncation cannot always be confirmed convincingly
because of the often small numbers of young clusters available. We can
gain statistical insights based on Chandar et al.'s (2010) M51 dataset
by conducting a Monte Carlo test to check for the existence, if any,
of a truncation mass in the clusters' MF (e.g., Bastian et
al. 2012). We derive $M_\star \approx 10^5 M_\odot$, with $\beta \sim
2$, both for cluster ages younger than $10^7$ yr and for ages up to
$10^8$ yr: see Fig. \ref{M51_CF_7_1} and \ref{M51_CF_8_1},
respectively. To our knowledge, this is the first confirmation of a
truncation mass for such young clusters in a spiral galaxy. We note
that Chandar et al. (2011) did not find any evidence for curvature in
the MF of their M51 cluster sample. We cannot directly compare our
results with theirs, however, since Chandar et al. (2011) analyzed the
MF for clusters with ages in the range of $(1-4) \times 10^8$ yr,
which is significantly older than the young age range considered
here. Those authors comment on the need to consider the effects of
cluster disruption, for which they quote a typical timescale of $2
\times 10^8$ yr for $10^4 M_\odot$ clusters. Asuming ``standard''
disruption analysis (see Chandar et al. 2011), our young sample is
expected to be negligibly affected by such effects.

We note that the fit result strongly depends on the lower mass limit
adopted. The lower mass limit is mainly affected by the level of the
sample's incompleteness. In turn, this dependence can be used, in
fact, to determine the statistically meaningful sampling limit to the
low-mass end of randomly sampled clusters. For a minumum mass of $10^3
M_\odot$ (following PA13), the dataset is poorly described by either a
pure power-law or a Schechter MF (see Fig. \ref{M51_CF_7_2}). The best
fit is obtained for a lower mass limit of approximately 5000--7000
$M_\odot$. As we will show in Section 5.2, our simulations based on
mock cluster populations imply that a completeness level $\gtrsim
80$\% would be suitable to construct cumulative MFs. We will also show
that the M51 cluster sample is best characterized using a completeness
limit in the range from $\sim 3500 M_\odot$ to $5000 M_\odot$.

In Fig. \ref{CF} we show the cumulative fraction of star cluster
masses in M51. The derived mass distributions for clusters with ages
younger than the upper limits indicated are shown as solid (red)
bullets. Monte Carlo simulations based on adoption of the same number
of clusters for a pure power-law distribution and a Schechter MF are
indicated by the blue and green lines, respectively, while the dashed
and dotted lines enclose 50\% and 90\% of the simulations,
respectively. It is clear that the M51 cluster MF is not well
approximated by a pure power law. Instead, the dataset requires a
truncation at the high-mass end. The truncation mass varies by a
factor of 1.5 for different ages, corresponding to $\Delta
\log(M_\star/M_\odot) \la 0.2$.

Bastian et al. (2012) also found that the M83 cluster MF is truncated
at the high-mass end, and that the truncation mass in the inner region
is 2--3 times higher than that in the outer region. Konstantopolous et
al. (2013) found an offset of $\sim 0.5$ dex (a factor of $\sim 3$)
between the inner disk and the outer regions of M83. The simulations
undertaken by both teams of the cluster mass distributions in the
galaxy's outer region show significant deviations from the ``best''
fit: see Fig. 16 of Bastian et al. (2012) and Fig. 15 of
Konstantopolous et al. (2013). The differences found for the inner and
outer subsamples in M51 derived here are significantly smaller, while
the MF fits are much better than those of either Bastian et al. (2012)
or Konstantopolous et al. (2013). The differences reported in the
literature correspond to $\Delta \log(M_\star/M_\odot) \sim
0.3$--0.45. Although a detailed re-analysis of the Bastian et
al. (2012) and Konstantopolous et al. (2013) results is beyond the
scope of this paper, we point out that such differences are often of
the same order of magnitude as the uncertainties in the individual
cluster masses (e.g., Anders et al. 2004; de Grijs et al. 2005). We
therefore urge the use of caution when interpreting differences of
this magnitude. Stochastic effects may also play a role in the
construction of the cumulative MFs routinely used in this field: a
single deviating cluster may skew the entire distribution, as can be
seen by a careful examination of a number of sudden ``jumps'' in the
cumulative MFs published to date as well as in this paper.

\section{Cluster Formation in M51}
\label{M51}

M51 is an interacting, grand-design spiral galaxy with a Seyfert 2
active galactic nucleus, projected on the sky in the constellation
Canes Venatici. We adopt inclination and position angles of
$20^{\circ}$ and $170^{\circ}$, respectively \citep{Tully1974}, and a
distance of $D = 8.4$ Mpc \citep{Feldmeier1997}.

We apply both maximum age and minimum mass cut-offs to our working
sample. There are two peaks in the age distribution at $\sim 5$ Myr
and $\sim 100$ Myr. We opted to impose a default maximum age of 10 Myr
for our initial analysis, following PA13, which includes 57\% of the
cluster population.

\begin{figure}[htbp]
    \centering
    \subfigure[Radius $< 4$ kpc]{
        \includegraphics[scale=0.35]{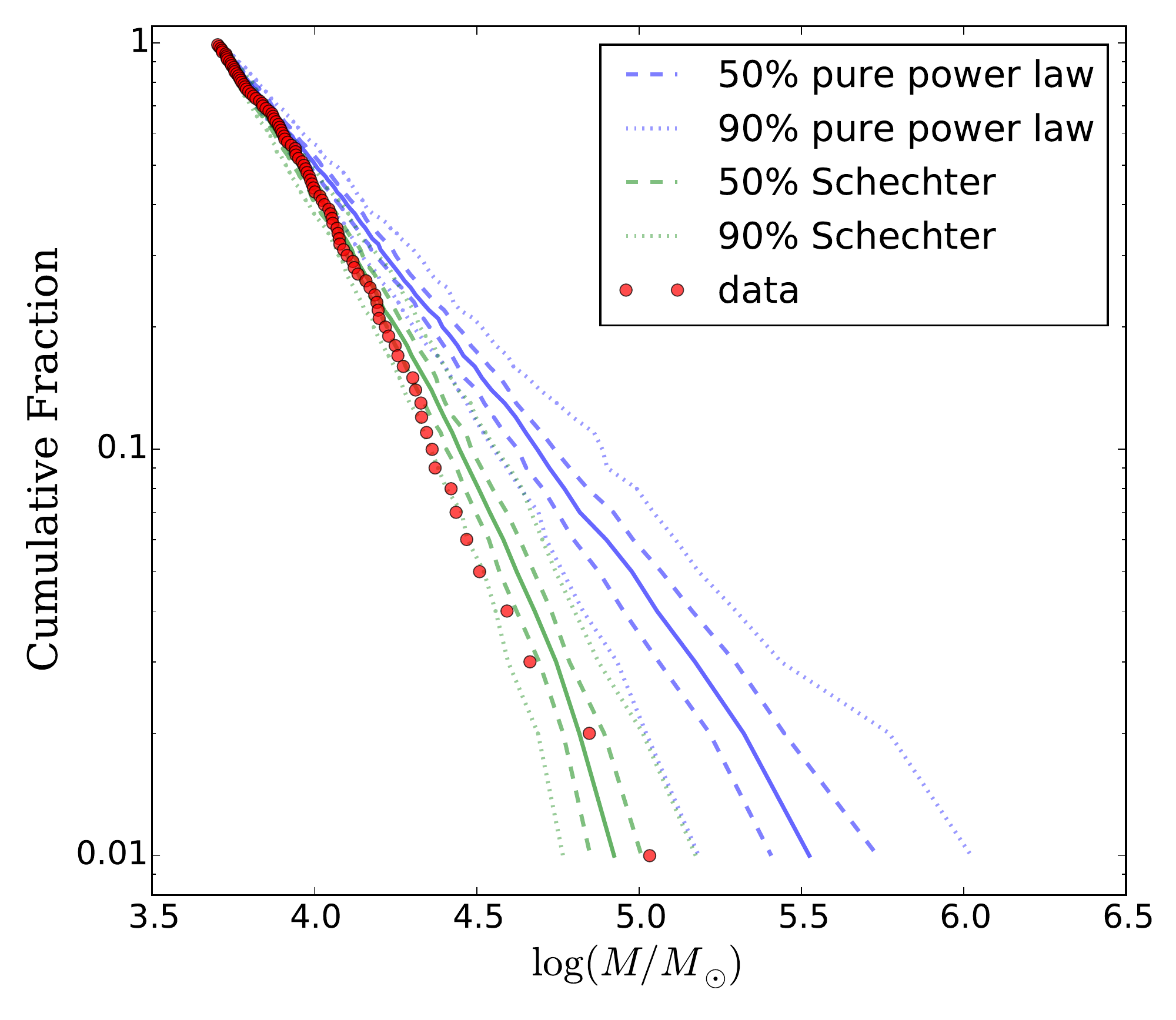}
        \label{M51_CF_7_in}
    }
    \subfigure[Radius $> 4$ kpc]{
        \includegraphics[scale=0.35]{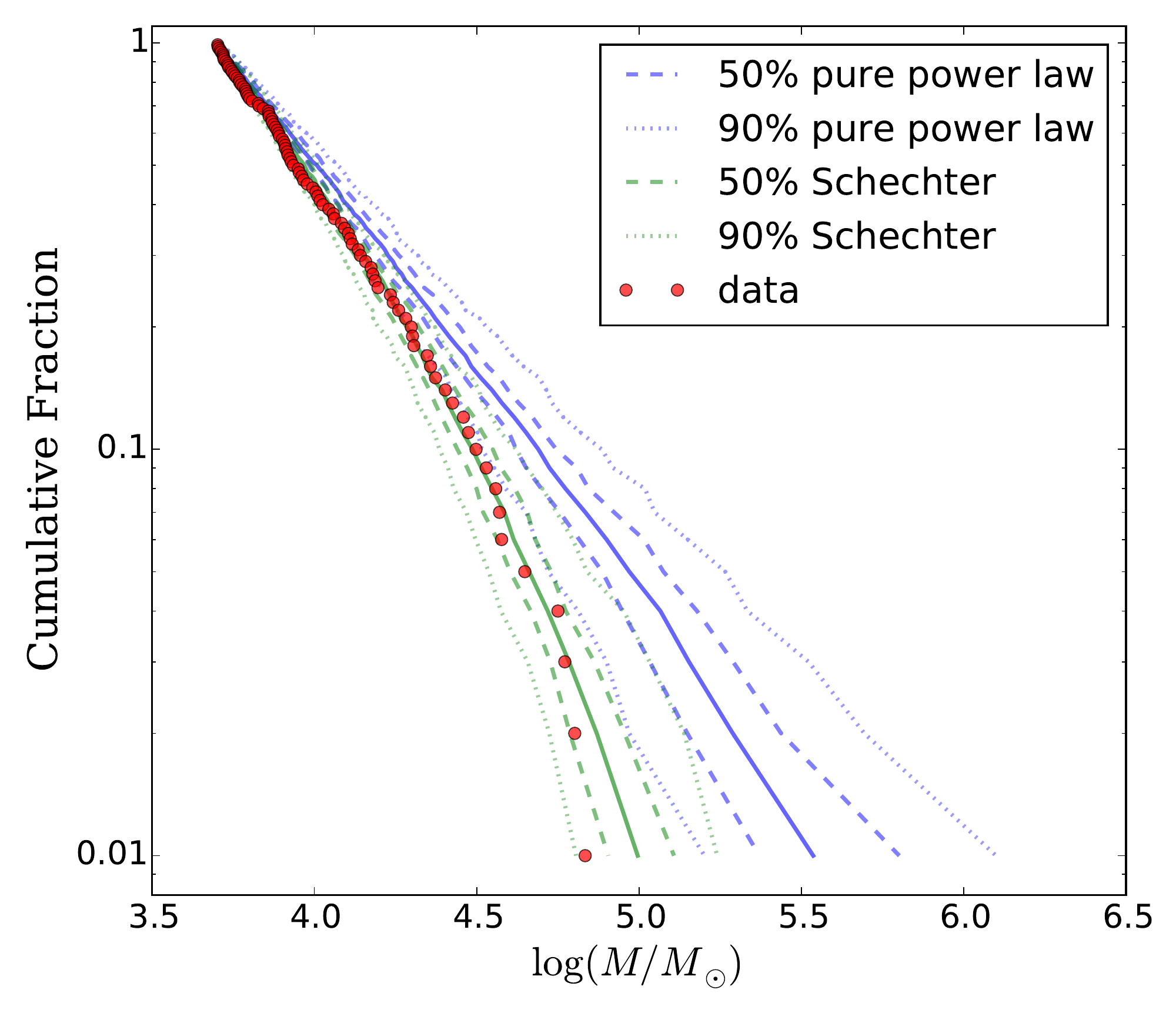}
        \label{M51_CF_7_out}
    }
    \caption{Cumulative cluster mass functions in the inner ($R_{\rm
        GC} <4$ kpc) and outer regions of M51. The legends are the
      same as in Fig. \ref{CF}. (a) $M_\text{min}=5000$ $M_\odot$,
      $\log(M_\star/M_\odot)=5$, $\beta=2$; (b) $M_\text{min}=5000$
      $M_\odot$, $\log(M_\star/M_\odot)=5.17$, $\beta=2$.}
    \label{CF_in_out}
\end{figure}

In Fig. \ref{CF_in_out} we show the best-fitting results for a set of
Monte Carlo simulations that were similarly stochastically sampled
from an underlying Schechter cluster MF as the observational data, for
a power-law index $\beta = -2$. We show the results for both the inner
and the outer regions of M51, adopting a radius of 4 kpc as our
operational boundary. The truncation mass was chosen in the same way
as for Fig. \ref{CF}: $M_\star = 1.0 \times 10^5$ $M_\odot$ and $1.5
\times 10^5$ $M_\odot$ for the inner and outer regions, respectively.

The masses of the most massive star clusters are well-determined,
since they are not strongly affected by stochastic IMF sampling
effects. However, because of stochastic IMF effects, the determination
of star cluster masses based on integrated photometry may become
highly uncertain for low-mass clusters \citep[e.g.,][and references
  therein]{Apellaniz2009, Fouesneau2010, Anders13, deGrijs2013}. We
thus also adopt a minimum cluster mass for our cluster sample
selection. The choice of the lower mass limit is important, because
the presence or absence of low-mass star clusters can affect the
distribution of the radial bins if the latter are selected to contain
constant numbers of clusters.

When fitting the data with a Schechter function, we adopt a minimum
mass of $M_{\rm cl} = 5000 M_\odot$. We conclude that reducing the
low-mass limit to values much below $5000 M_\odot$ leads to
significant instability in the fits and this thus affects the eventual
robustness of the parameters returned by the fits. Application of this
selection limit ensures that we retain 17\% of the total sample of
cluster candidates (633 objects). Our simulations result in better
fits than those previously obtained by both Bastian et al. (2012) and
Konstantopolous et al. (2013). By comparison with the latter authors,
it appears that this difference is largely caused by the differences
in the adopted mass limits.

\begin{figure}[htbp]
  \centering
  \subfigure[]{
  \includegraphics[scale=0.25]{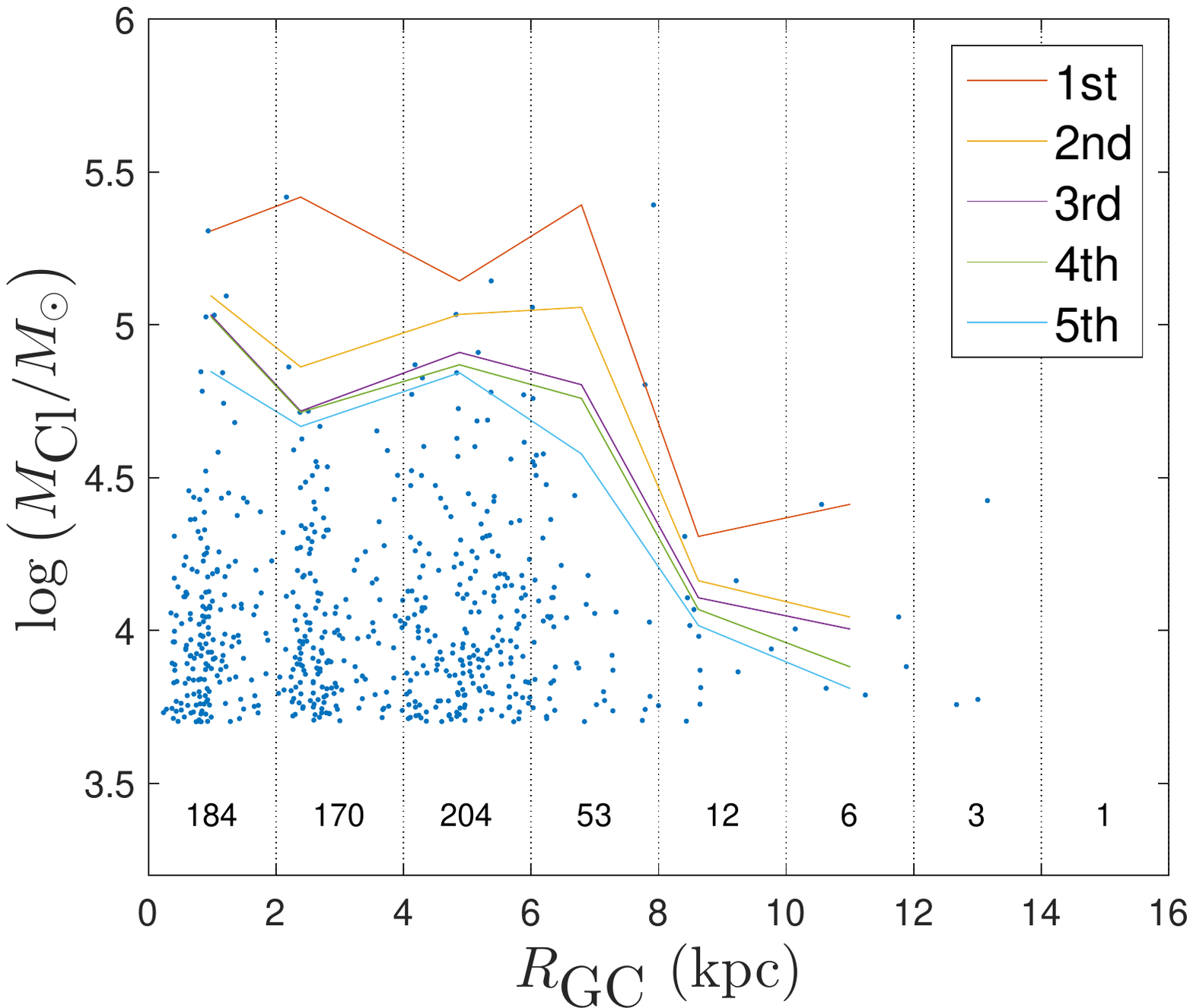}
  }
  \subfigure[]{
  \includegraphics[scale=0.25]{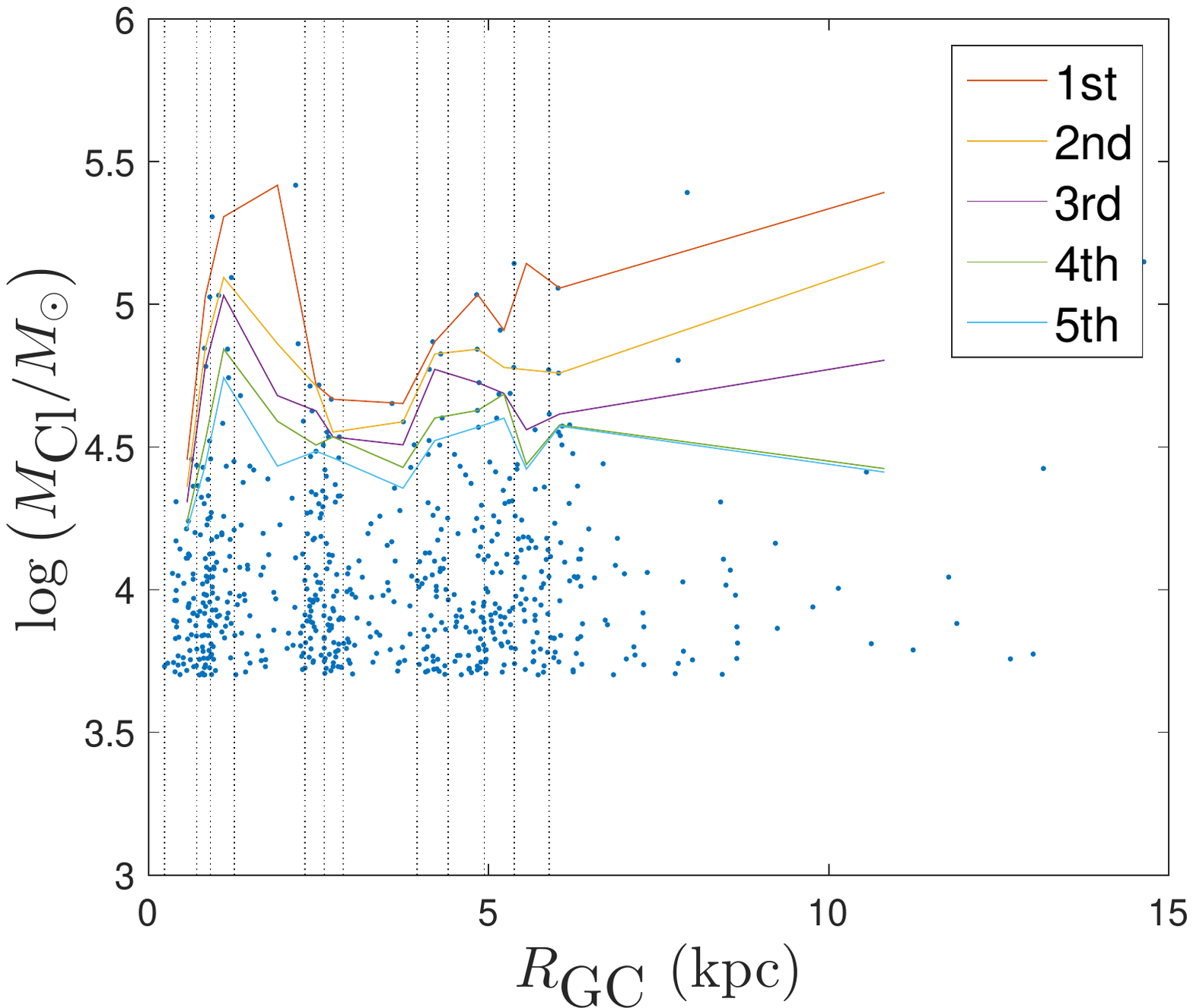}
  }
  \\
  \subfigure[]{
  	\includegraphics[scale=0.25]{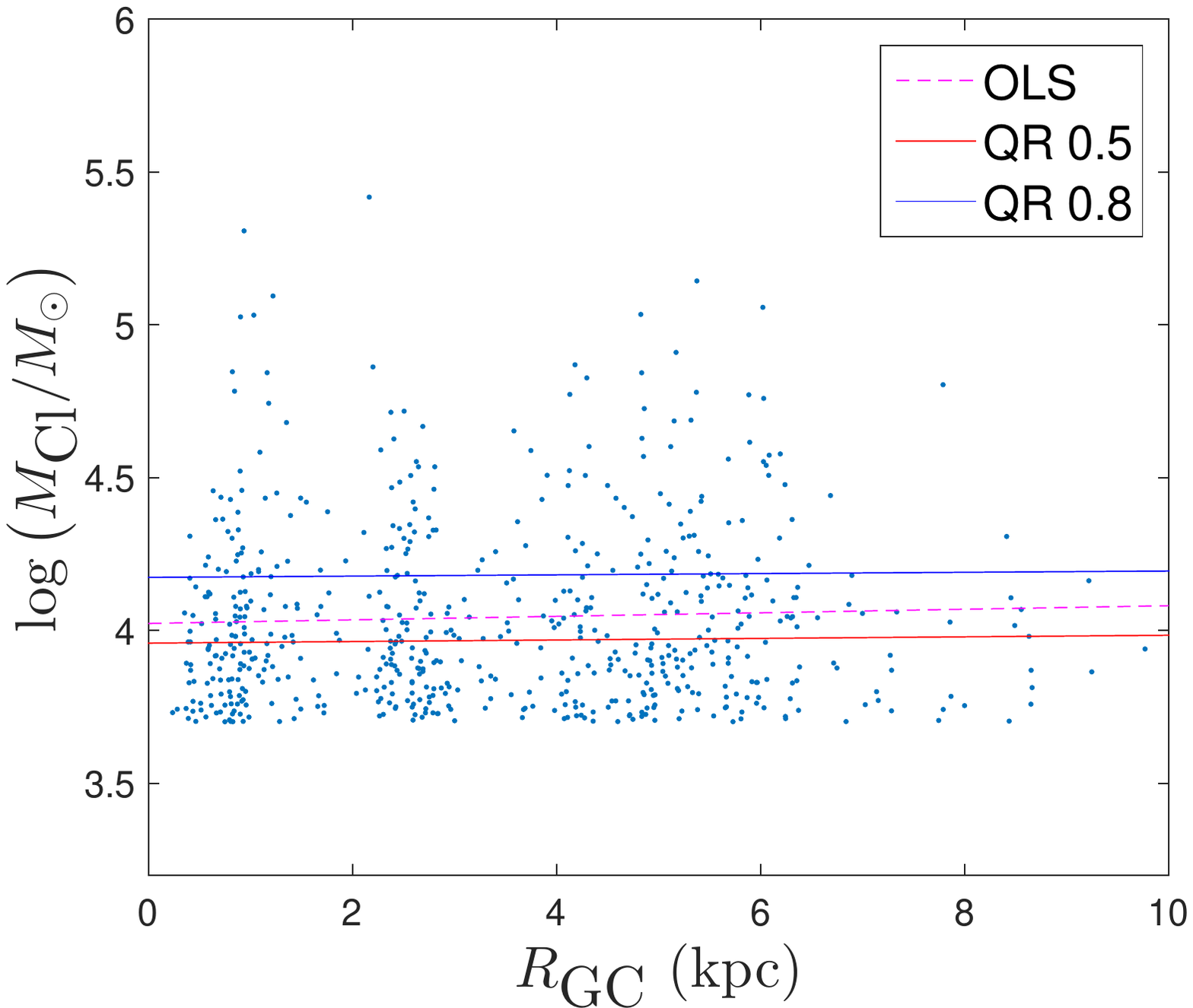}
  }
  \subfigure[]{
	\includegraphics[scale=0.25]{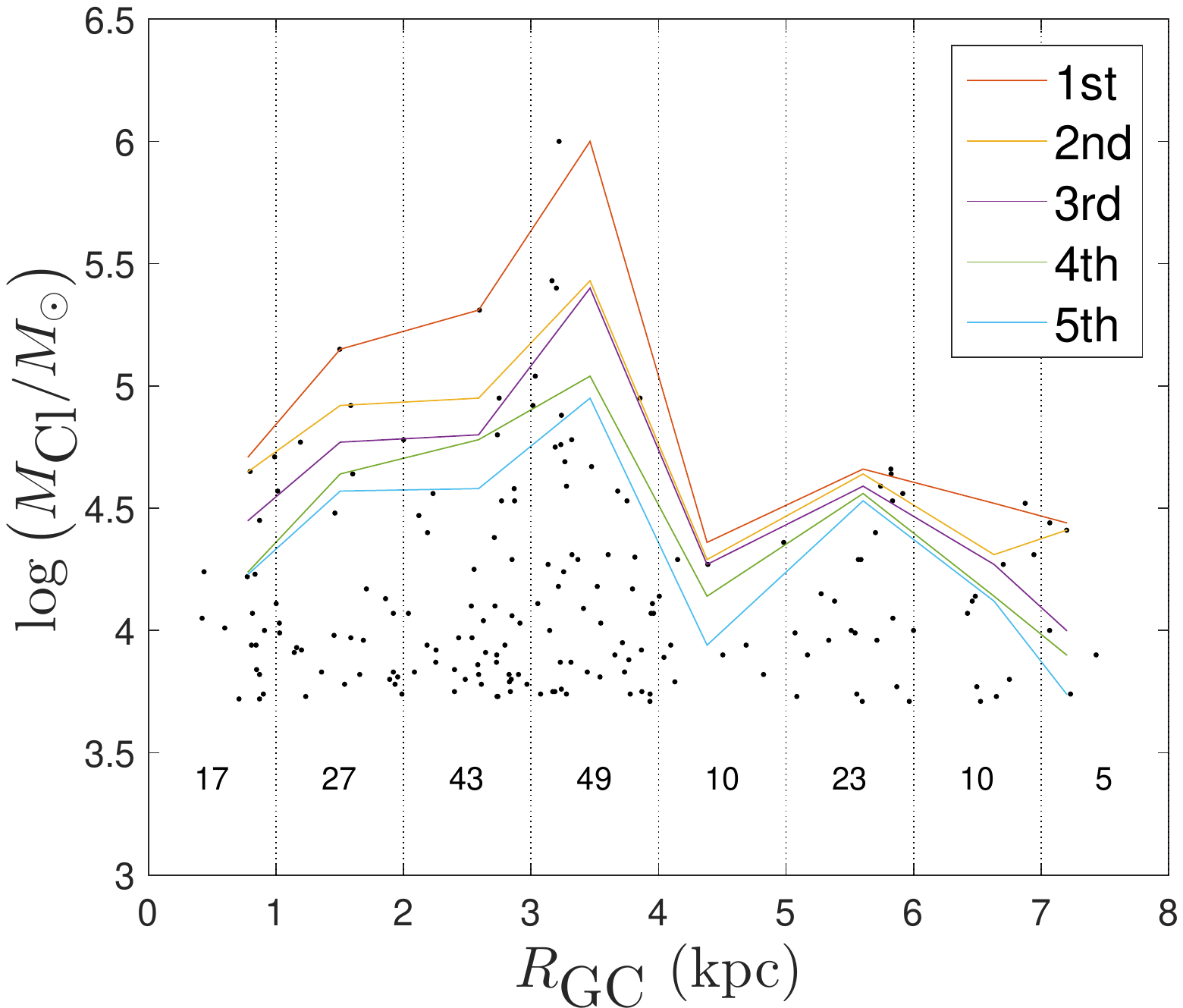}
  }
\caption{(a) Young M51 star cluster masses as a function of
  galactocentric radius, adopting a minimum mass limit of $M_{\rm cl}
  = 10^3 M_\odot$. The radial bin size adopted is 2 kpc; the numbers
  shown in each bin indicate the cluster numbers included. The solid
  lines connect the $i^{\rm th} (i = 1, 2, 3, 4, 5)$ ranked most
  massive clusters in each bin. The effects of varying the bin sizes
  will be explored in Section 4.3. (b) As (a), but using radial bin
  sizes containing constant cluster numbers ($N = 50$), as indicated
  by the vertical dashed lines. (c) Comparison of the results of our
  linear OLS fit (purple dashed line) with those resulting from QR for
  $\tau = 0.5$ (red solid line) and $\tau = 0.8$ (blue solid line),
  where $\tau$ refers to the $\tau^{\rm th}$ sample quartile; see
  Section 4. (d) As (a) but for the M83 star cluster system, adopting
  radial bin sizes of 1 kpc.}
\label{Mass_seq_All}
\end{figure}

Figure \ref{Mass_seq_All}a shows the distribution of the young cluster
masses in M51 as a function of galactocentric radius, where the age
and mass ranges have been restricted as discussed above. The clusters
are grouped here in radial bins with fixed widths of 2 kpc each. The
differently colored solid lines connect the $i^{\rm th} (i = 1, 2, 3,
4, 5)$ ranked most massive clusters in each bin. The numbers at the
bottom of each bin indicate the number of young clusters contained in
the relevant bin.

There is a clear general tendency that the mass of the $i^{\rm th}$
ranked most massive cluster decreases with increasing galactocentric
distance, at least out to $R_{\rm GC} \simeq 7$--8 kpc. However, the
number of clusters in each bin also decreases, which signals that the
observed trend may be confounded by size-of-sample effects. PA13
suggested that, at first glance, a similar result they obtained for
their M33 objects may correspond to the simple, universal probability
density distribution function (PDF) given by the ICMF. However, they
pointed out that they cannot rule out a different model where the
formation of the most massive star clusters is limited by the local
physical conditions, e.g., by the local gas density (e.g.,
Pflamm-Altenburg \& Kroupa 2008).

To further examine the physical relevance of this possibility, we
tried to eliminate the size-of-sample effect by adopting radial bin
sizes such that each bin contained the same number of clusters. If the
ICMF represents a simple, universal, and environment-independent PDF,
then we would not expect to find any radial dependence of the mass of
the $i^{\rm th}$ ranked most massive star cluster. We adopted bins
containing 60 clusters each for our analysis: see
Fig. \ref{Mass_seq_All}b. In each bin, we identified the $i^{\rm th}$
ranked most massive young star clusters. The galactocentric radii
assigned to the $i^{\rm th}$ ranked most massive star clusters are the
averages of the radial distances of all star clusters in the relevant
bin.

We first used the OLS method to fit straight lines to the data,
\begin{equation}
\log(M_{\text{cl}}/{M_{\odot}}) = a R_{\rm GC} ({\text{kpc}}) + b,
\end{equation}
where $a$ and $b$ are the slope and intercept, respectively. Our fit
results for M51 are rather different from those of PA13 for M33. The
latter authors found a maximum absolute value of the slope for all
clusters of 0.158 (for all data sets considering the second to the
fifth most massive clusters in each bin), while for the same relative
masses our maximum absolute value for the slope is merely 0.036. This
difference has important implications for our understanding of the
cluster-formation mode in the host galaxy. If the ICMF would be
independent of its natal environment, i.e., if cluster formation were
to proceed stochastically, the expected slope for the young massive
clusters would be zero, modulo the size-of-sample effect. The observed
value of the slope of the cluster mass--galactocentric radius relation
thus becomes a useful criterion to test statistically for ICMF
departures from the hypothesis of a universal (``stationary,'' in
statistical terms) stochastic cluster-formation process.

\begin{table}[htbp]
\caption{Slopes of the radial distribution of the $i^{\rm th}$ ranked
  most massive star cluster in each bin in M51. The conditions imposed
  on our datasets are discussed in Section \ref{Parameter}.}
\label{tbl_M51}
\tiny
\begin{center}
\begin{tabular}{ccccc}
\tableline\tableline
Condition                                  & $i^{\rm th}$ & Slope    & Intercept & $p$ value \\
                                           &              & ($a$)    & ($b$)     &         \\
\tableline
\multirow{5}{*}{}
$N = 50$                                   & 1            & 0.045    & 4.8       & 0.26    \\
$\log\left( M_{\rm{cl}} / M_\odot \right)$ & 2            & 0.028    & 4.7       & 0.081   \\
$> 3.7$;                                   & 3            & 0.0060   & 4.6       & 0.80    \\
$\log(t \mbox{ yr}^{-1})$                  & 4            & $-$0.0051& 4.6       & 0.71    \\
$\le 7$                                    & 5            & 0.0042   & 4.5       & 0.77    \\
\tableline
$N = 50$                                   & 1            & 0.046    & 4.8       & 0.21    \\
$\log\left( M_{\rm{cl}} / M_\odot \right)$ & 2            & 0.031    & 4.6       & 0.056   \\
$> 3.7$;                                   & 3            & 0.028    & 4.5       & 0.24    \\
$\log(t \mbox{ yr}^{-1})$                  & 4            & $-$0.0041& 4.6       & 0.81    \\
$\le 7.3$                                  & 5            & 0.00091  & 4.5       & 0.96    \\
\tableline
$N = 50$                                   & 1            & 0.025    & 4.8       & 0.48    \\
$\log\left( M_{\rm{cl}} / M_\odot \right)$ & 2            & 0.010    & 4.6       & 0.70    \\
$> 3.5$;                                   & 3            & $-$0.0025& 4.5       & 0.89    \\
$\log(t \mbox{ yr}^{-1})$                  & 4            & 0.0041   & 4.4       & 0.81    \\
$\le 7$                                    & 5            & 0.00067  & 4.4       & 0.97    \\
\tableline
$N = 30$                                   & 1            & 0.037    & 4.7       & 0.29    \\
$\log\left( M_{\rm{cl}} / M_\odot \right)$ & 2            & 0.0032   & 4.6       & 0.90    \\
$> 3.7$;                                   & 3            & 0.0078   & 4.5       & 0.65    \\
$\log(t \mbox{ yr}^{-1})$                  & 4            & 0.00094  & 4.4       & 0.95    \\
$\le 7$                                    & 5            & $-$0.0021& 4.4       & 0.89    \\
\tableline
$N = 40$                                   & 1            & 0.037    & 4.8       & 0.30    \\
$\log\left( M_{\rm{cl}} / M_\odot \right)$ & 2            & 0.024    & 4.6       & 0.15    \\
$> 3.7$;                                   & 3            & 0.036    & 4.5       & 0.050   \\
$\log(t \mbox{ yr}^{-1})$                  & 4            & 0.0033   & 4.5       & 0.75    \\
$\le 7$                                    & 5            & 0.011    & 4.4       & 0.34    \\
\tableline
$N = 60$                                   & 1            & 0.043    & 4.8       & 0.34    \\
$\log\left( M_{\rm{cl}} / M_\odot \right)$ & 2            & 0.025    & 4.7       & 0.17    \\
$> 3.7$;                                   & 3            & 0.020    & 4.6       & 0.39    \\
$\log(t \mbox{ yr}^{-1})$                  & 4            & $-$0.012 & 4.6       & 0.28    \\
$\le 7$                                    & 5            & $-$0.004 & 4.5       & 0.74    \\
\tableline
\end{tabular}
\end{center}
\end{table}

To gain a better insight into the cluster mass--galactocentric radius
relation characteristic of the M51 cluster population, we next used
the QR method (Koenker 2005; Feigelson \& Babu 2012; Ivezi\'c et
al. 2014). QR analysis is often used in statistics and
econometrics. Whereas the OLS method results in estimates that
approximate the conditional mean of the response variable (in the
context of the analysis of this paper, the response variable is the
cluster mass) given certain values of the predictor variables (here,
the galactocentric distances), QR aims at estimating either the
conditional median or other quantiles of the response variable. QR
estimates are more robust against outliers in the response
measurements than their OLS-based counterparts. In addition, QR
analysis does not impose arbitrary binning, such as that required
above to bend OLS to the purpose of quantile fitting.

Symmetry implies that the minimization of the sum of the absolute
residuals must be equal to the number of positive and negative
residuals. Since the symmetry of the absolute values yields the
median, minimizing a sum of asymmetrically weighted absolute
residuals---i.e., simply giving differing weights to positive and
negative residuals---yields the quantiles. We thus need to solve
\begin{equation}
\min \limits_{\xi\in\Re}\sum \rho_\tau \left( y_i-\xi \right),
\end{equation}
where the function $\rho_\tau(\cdot)$ is the tilted absolute value
function targeting the $\tau^{\rm th}$ sample quantile. To model a
dependence of the median on a given variable and thus obtain an
estimate of the conditional median function, we need to replace the
scalar $\xi$ by the parametric function $\xi(x_i,\beta)$ and adopt
$\tau = \frac{1}{2}$. To obtain estimates of the other conditional
quantile functions, we replace the absolute values by
$\rho_\tau(\cdot)$ and solve
\begin{equation}
\min \limits_{\xi\in\Re p}\sum \rho_\tau \left( y_i-\xi(x_i,\beta)
\right).
\end{equation}
When $\xi(x_i,\beta)$ is a linear function of parameters, the
resulting minimization problem can be solved very efficiently using
linear programming methods.

QR examines the relation between the response variable (cluster mass)
and predictor variable (galactocentric distance) for a quantile
$\tau$. Unlike OLS regression, we now have a family of curves to
interpret, and we can focus our attention on the particular segments
of the conditional distribution, thus obtaining a more complete view
of the relationship between the variables
\citep{Koenker2001}. Although we employ a frequentist QR approach in
the present work (adapted from Koenker 2009), it is worth noting that
Bayesian versions exist (e.g., Yu \& Moyeed 2001), which can be
readily incorporated into hierarchical models to account for
complexities, such as the presence of large uncertainties in the
explanatory variables.

We selected the M51 cluster sample from Fig. 1a and 1b to demonstrate
the QR result. The data selection in Fig. \ref{Mass_seq_All}c is
identical to that used for the OLS regression. The red dashed line
represents the linear OLS fit, while the blue and black solid lines
represent the QR results for $\tau = 0.5$ and $\tau = 0.8$,
respectively. The line pertaining to the QR result for $\tau=0.5$ has
a similar intercept and slope as that resulting from the OLS
regression, whereas QR for $\tau=0.8$ yields a much larger intercept,
although it returns a similar slope.

Figure \ref{Age_seq} shows the slopes as a function of $\tau$ for
different age ranges. Except for the smallest regressed quantiles
(where a null slope is enforced by our fixed lower mass limit), the
corresponding galactocentric radius is associated, for the maximum
likelihood solution, with a very small decrease in the mass
distribution. Taking the standard error ($\sigma$) into account, a
Wald test (Wald 1939) rules against rejection of the null hypothesis
(zero slope) for all quantiles. This is so, because the approximate
$3\sigma$ confidence intervals encompass a slope of zero for (almost)
all quantiles. The simple frequentist $p$-value interpretation of the
Wald test then automatically recommends against rejecting the null
hypothesis. We point out that for a very small number of values of
$\tau$, and depending on the age range considered, the Wald test
cannot be applied directly. From a statistical point of view, this can
be understood by realizing that the Wald test is, strictly speaking,
only applicable to single quantiles, while in this case we run into
non-independent multiple hypothesis testing if we are to look over all
possible quantiles.

\begin{figure}[htbp]
	\centering
	\subfigure[]{
		\includegraphics[scale=0.35]{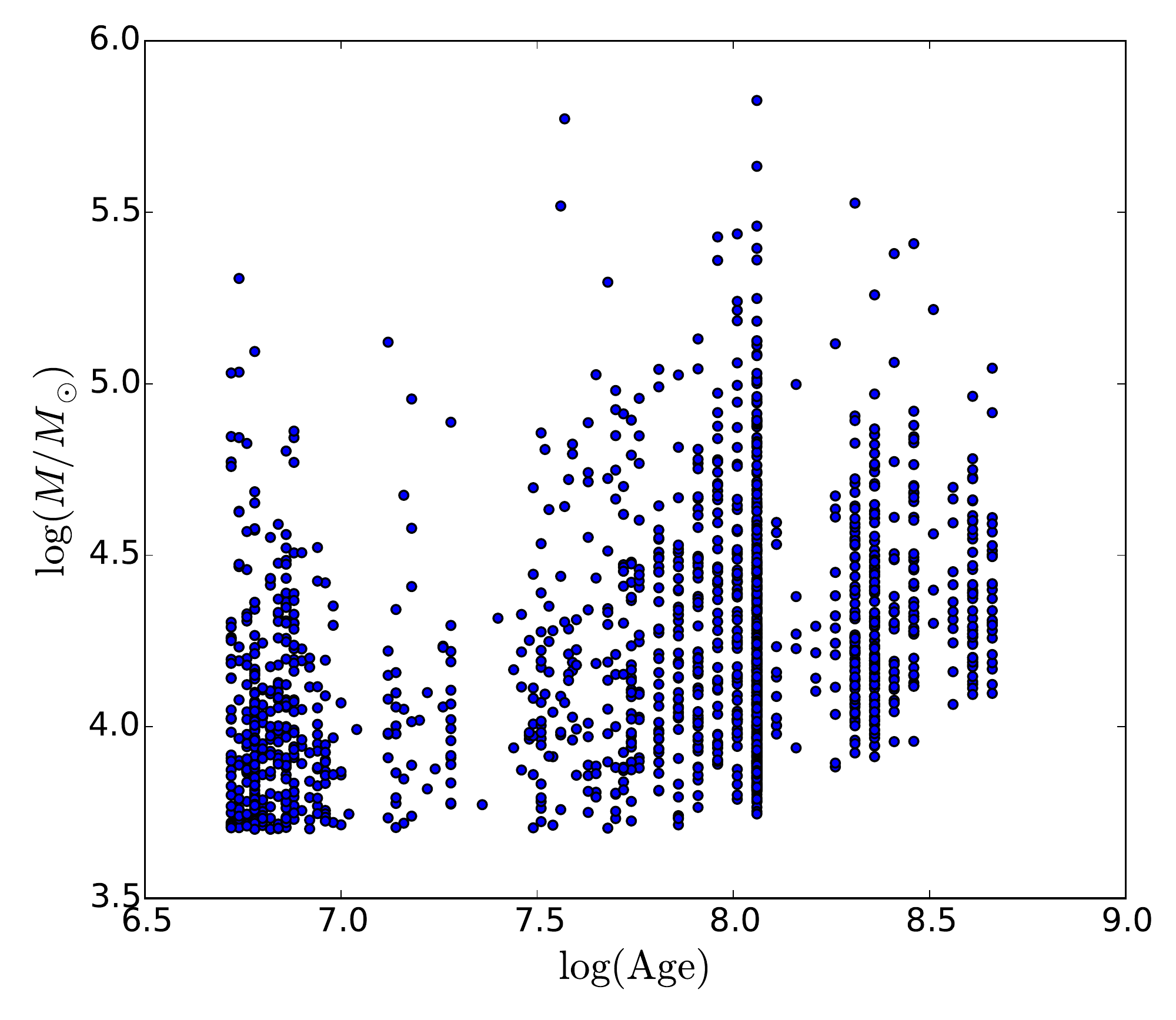}
	}
	\subfigure[]{
		\includegraphics[scale=0.35]{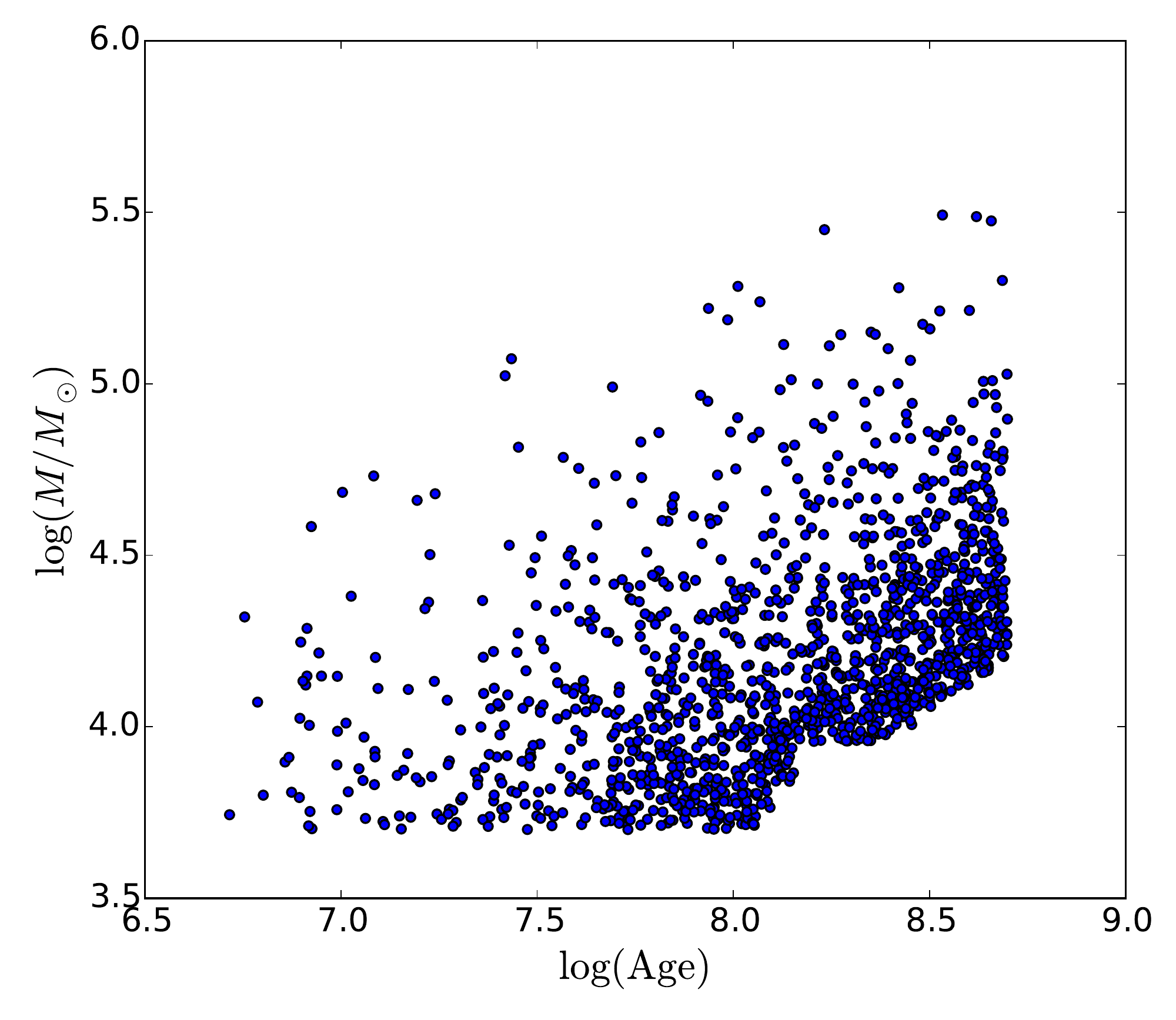}
	}
	\\
	\subfigure[]{
		\includegraphics[scale=0.35]{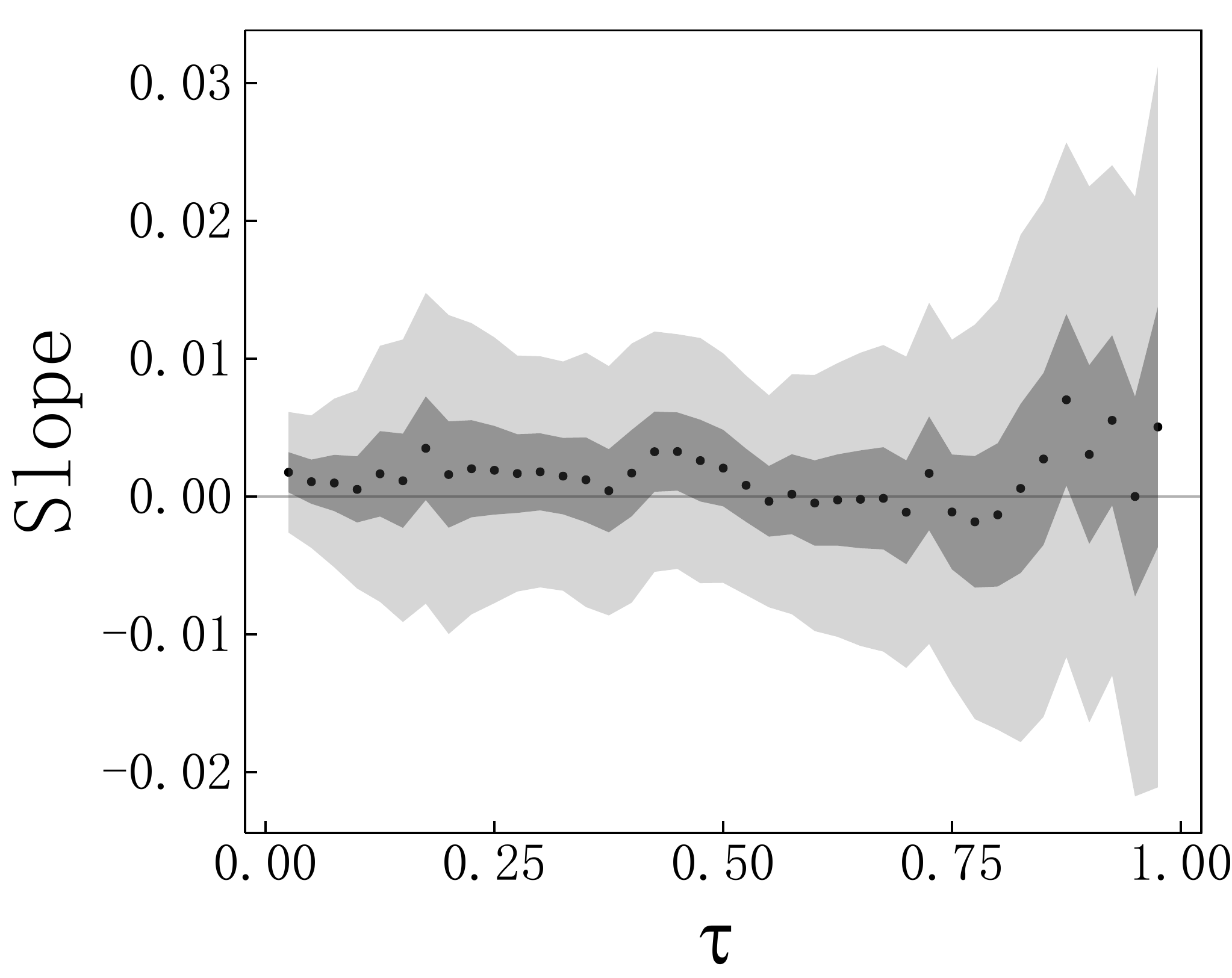}
	}
	\subfigure[]{
		\includegraphics[scale=0.35]{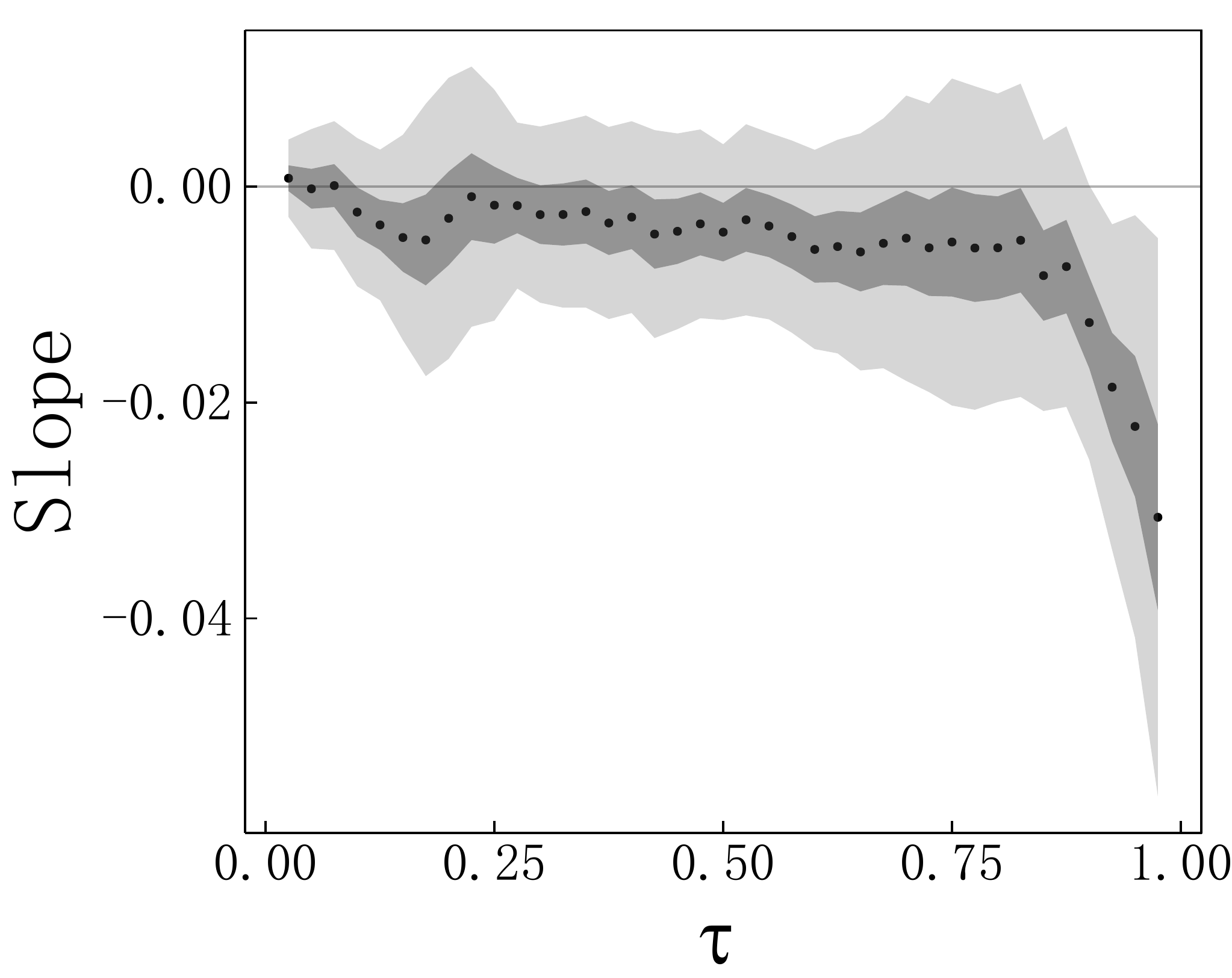}
	}
	\caption{(a) Age--mass distribution of our observational M51
          cluster sample. (b) Equivalent distribution of the simulated
          clusters. The magnitude limit adopted determines the number
          of sample clusters, while the total number of clusters above
          the adopted completeness limit is based on the number of
          clusters in our M51 sample. (c) QR result for clusters with
          a universal (radius-independent) truncation mass. (d) QR
          result for clusters characterized by a radius-dependent
          truncation mass.}
	\label{age_mass}
\end{figure}

Finally, to validate our observational results for M51, we constructed
a mock cluster population characterized by a Schechter-type MF. Using
our simulated cluster population, we explored two conditions, one
where the truncation mass does not depend on galactocentric distance
and a second where the truncation in the galaxy's inner regions occurs
for cluster masses that are 2--3 times higher than those in the
galactic periphery. We randomly assigned cluster ages from 5 Myr (the
youngest cluster age in our observational sample) to 500 Myr, adopting
a constant cluster formation rate for the entire intervening period
for simplicity. We used the Bruzual \& Charlot (2003) SSP models to
assign $B$-band magnitudes to our sample clusters, a choice driven by
the characteristics of the data set of Chandar et al. (2011), whose
cluster photometry is ultimately limited by their $B$-band
observations. We applied the observational magnitude limit from
Chandar et al. (2011) and adjusted the number of clusters to reproduce
the observational data set as closely as possible given the adopted
constraints. The distributions in the diagnostic age--mass plance of
both the observed M51 cluster population and our mock sample are shown
in Fig. \ref{age_mass}a and b.

First, we considered the cluster MFs for age ranges younger and older
than $10^8$ yr for a universal truncation mass, $M_\star = 10^5
M_\odot$. For the younger clusters, we adopted a low-mass limit of
$5000 M_\odot$. The resulting value of $M_\star$ was 50\% higher than
expected, $M_\star \simeq 2 \times 10^5 M_\odot$. For the older
cluster sample, with ages between 100 Myr and 500 Myr, we adopted a
low-mass limit of $10^4 M_\odot$, which resulted in a derived
truncation mass of $M_\star \simeq 1.25 \times 10^5 M_\odot$. This
difference in the derived truncation masses reflects the uncertainties
pertaining to our calculations. The corresponding QR test for this
mock data set is similar to what we found previously, resulting in a
close-to-zero slope for all quantiles.

Second, we considered the case of a truncation mass that depends on
galactocentric distance. Observationally, a galactocentric-distance
dependence has not been established for the M51 cluster
population. Therefore, we adopted Bastian et al.'s (2012) suggested
radial dependence for the M83 clusters for our mock test. They
reported truncation masses of $M_\star = 2 \times 10^5 M_\odot$ and
$M_\star = 10^5 M_\odot$ for, respectively, the inner and outer
regions of the galaxy. Applying our approach to the entire cluster
sample, we derive $M_\star \simeq 2 \times 10^5 M_\odot$, since the
number of clusters in the galaxy's inner regions is much larger than
that at larger radii. However, considering the truncation masses for
different radial ranges separately, we derived values consistent with
the theoretical input values, within the method's typical
uncertainties. Thus, if the truncation mass varies systematically by a
factor of 2 (or more) radially across a given galaxy, we should expect
this to show up in our results. The corresponding QR test also allows
us to distinguish between cluster MFs with truncation masses that
differ by at least a factor of 2.

\section{Parameter Dependence}
\label{Parameter}

In order to explore how the results depend on the adopted ICMF
parameters, we varied the maximum age and minimum mass limits, as well
as the radial bin size, and re-analyzed the clusters'
mass--galactocentric radius relationship. The maximum age limit
adopted determines the number of clusters included in our sample;
varying the maximum age limit leads to significantly different sample
sizes. The minimum mass limit adopted determines the distribution of
the star clusters in radial bins containing constant cluster
numbers. Varying the minimum mass limit will thus change the mass and
location of the most massive star cluster in each bin, which will
affect our analysis pertaining to stochasticity in the ICMF. Bin-size
variations also have an immediate effect on the resulting ICMF.

\subsection{Maximum Age Limit}

If we adopt a maximum age for the young clusters of $10^{7.3}$ yr (20
Myr), the resulting M51 cluster sample includes 689 objects. The
previously noted tendency that the maximum cluster masses decrease
with increasing galactocentric radius is still obvious. However, for
both the OLS (adopting equal-number radial bins) and QR analyses, the
resulting slopes of the linear fits tend to values close to zero. The
detailed results are included in Table \ref{tbl_M51}. The maximum
slope for the second or third most massive cluster in each bin ranges
from 0.025 to 0.036; for the first-ranked cluster in each bin, which
we treat separately for stochastic reasons, the maximum slope attained
ranges from 0.037 to 0.046.

\begin{figure}
	\centering
	\subfigure[$\log(t \mbox{ yr}^{-1}) \le 6.9$]{
	  	\includegraphics[scale=0.24]{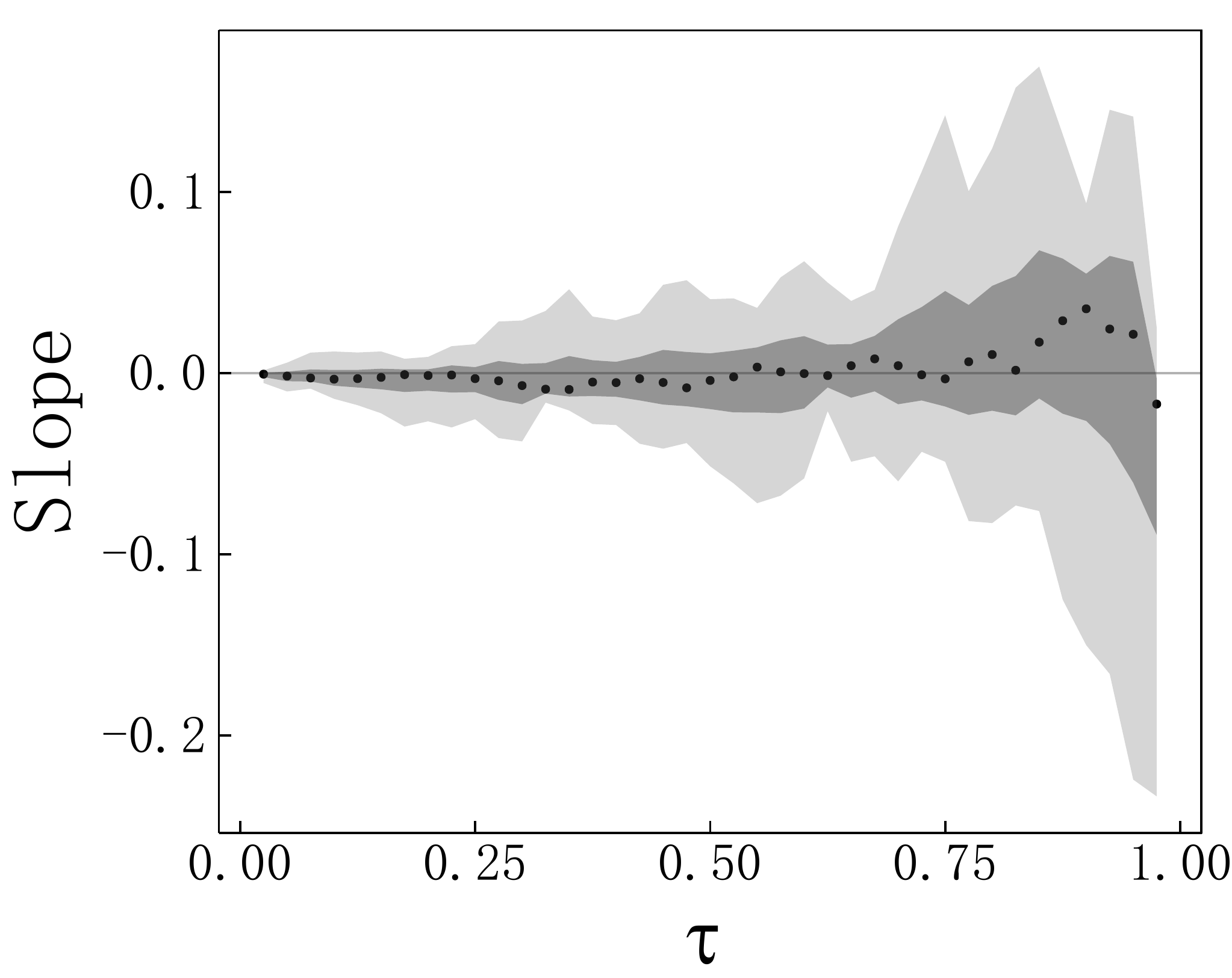}
	}
    \subfigure[$\log(t \mbox{ yr}^{-1}) \le 7.0$]{
    	\includegraphics[scale=0.24]{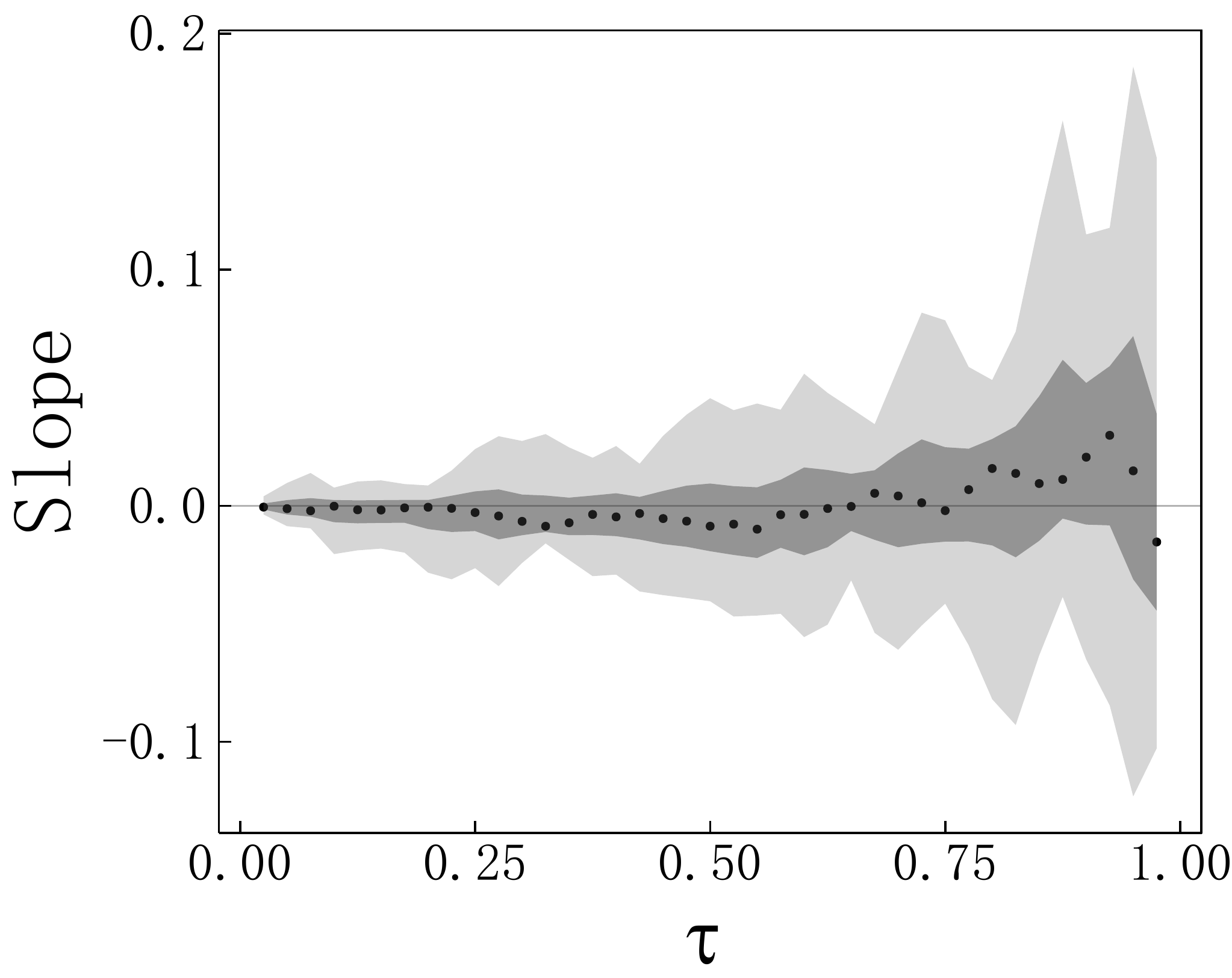}
    }
    \subfigure[$\log(t \mbox{ yr}^{-1}) \le 7.2$]{
      	\includegraphics[scale=0.24]{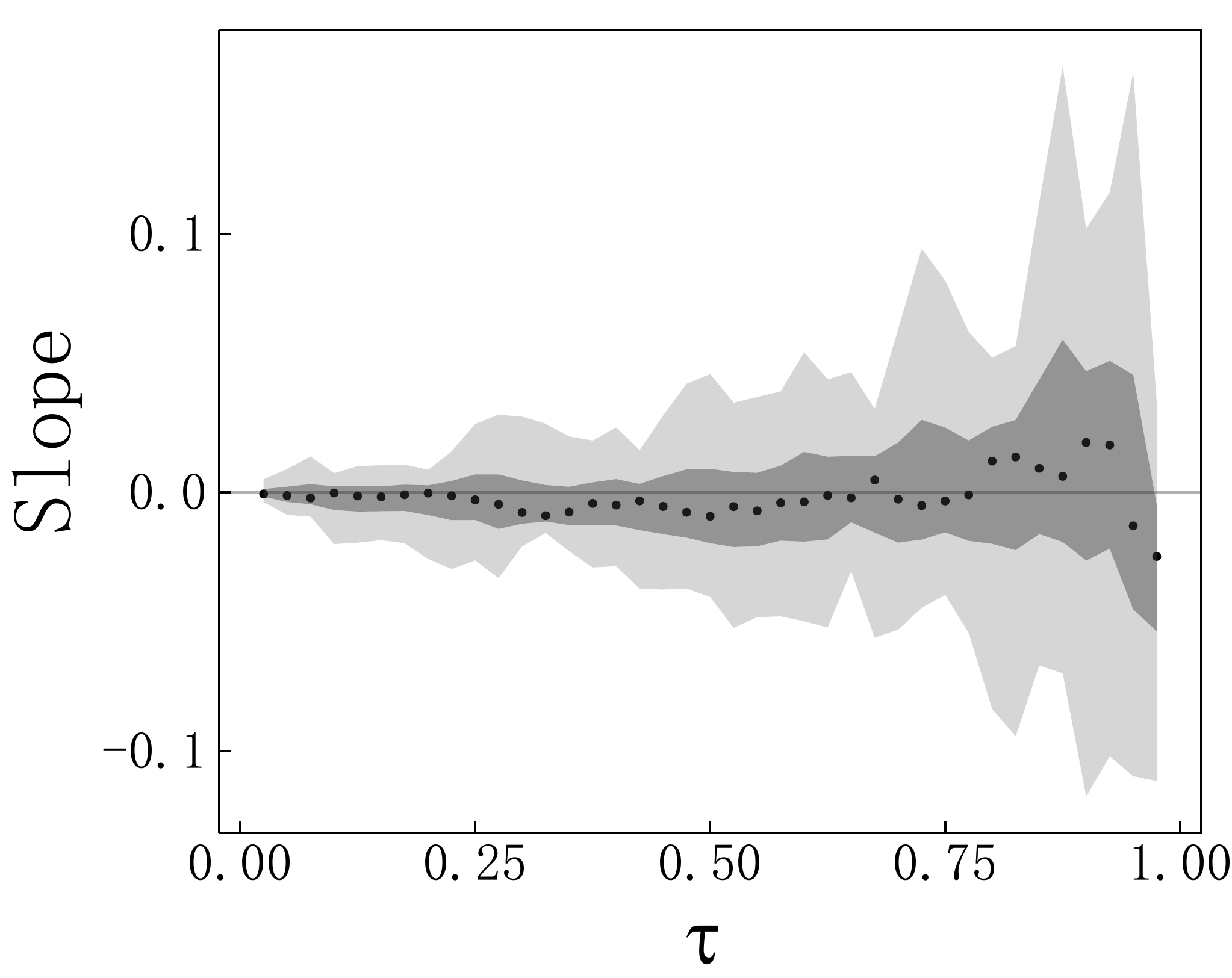}
    }
    \\
    \subfigure[$\log(t \mbox{ yr}^{-1}) \le 7.4$]{
    	\includegraphics[scale=0.24]{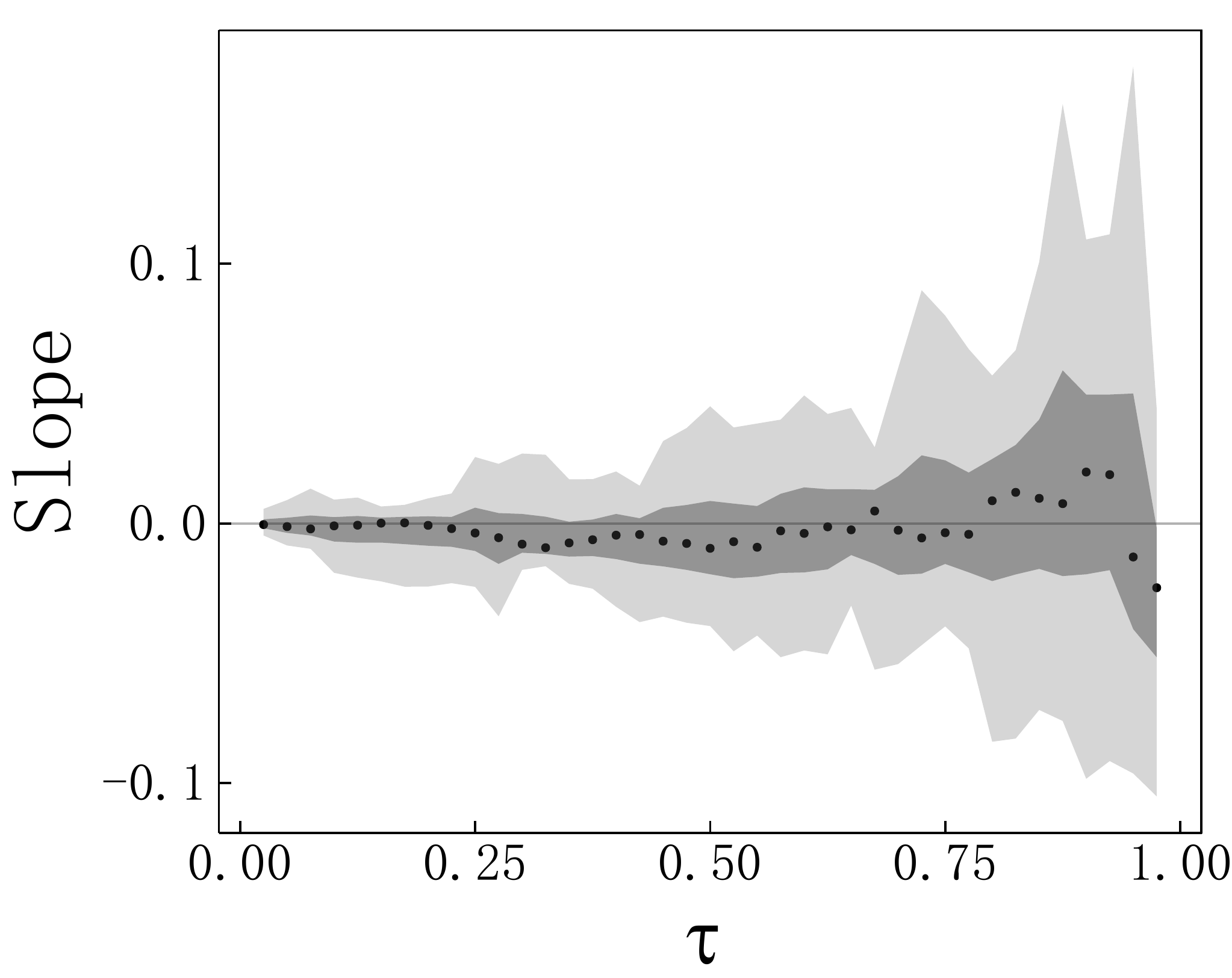}
    }
    \subfigure[$\log(t \mbox{ yr}^{-1}) \le 7.6$]{
      	\includegraphics[scale=0.24]{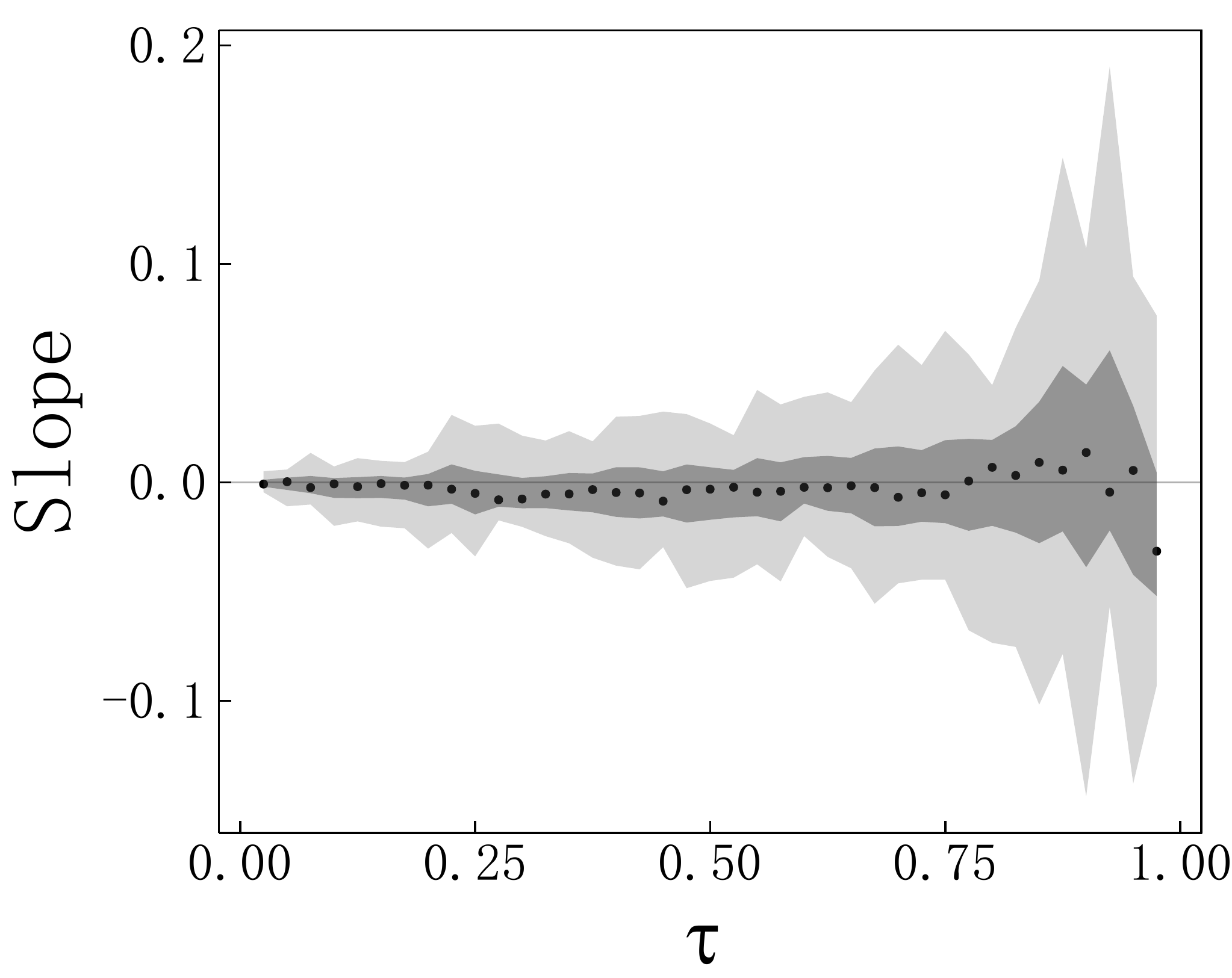}
    }  
    \subfigure[$\log(t \mbox{ yr}^{-1}) \le 7.8$]{
	    \includegraphics[scale=0.24]{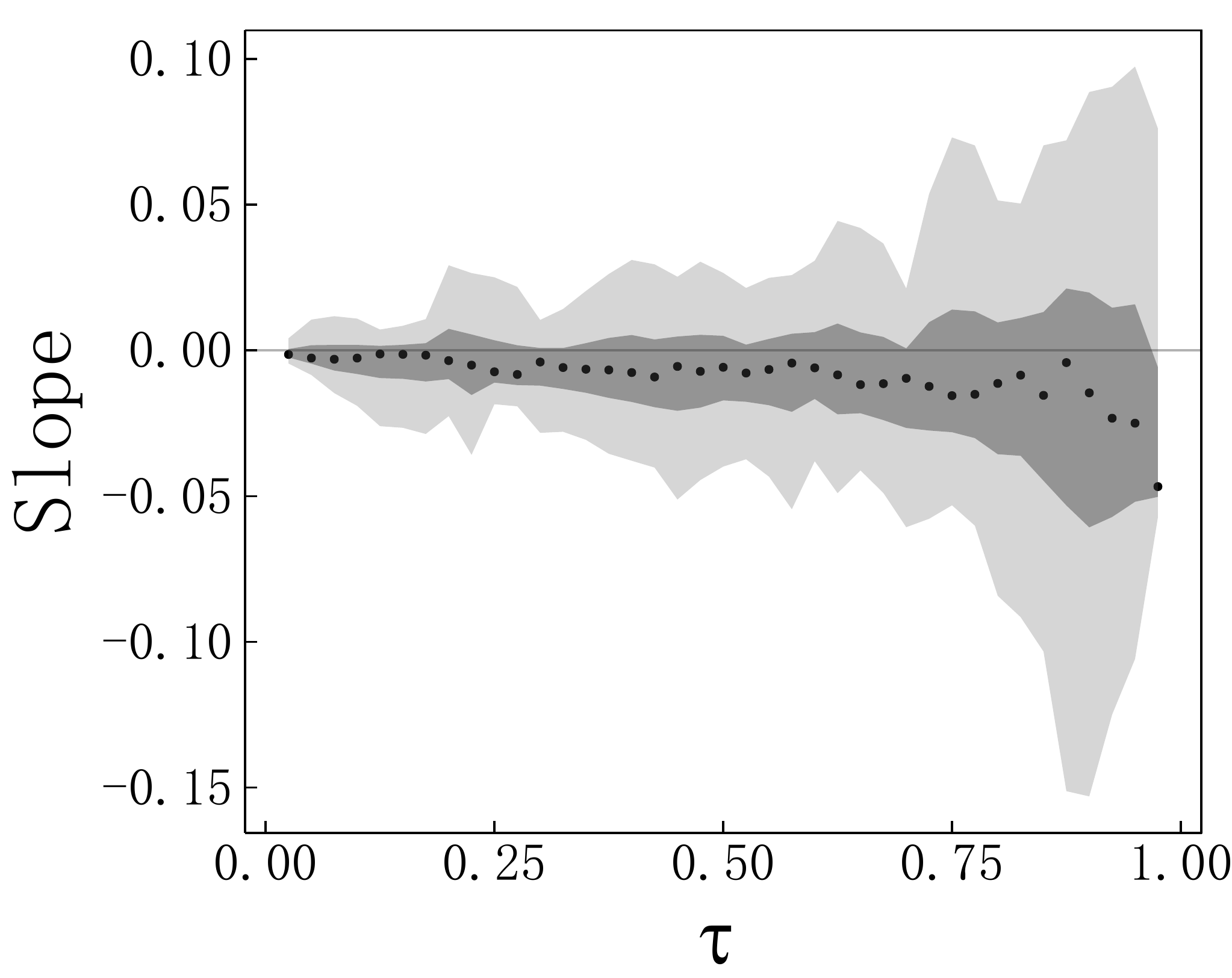}
    }
	\caption{QR results for different maximum age limits in M51,
          as indicated. The slope of the estimated linear QR is
          plotted as a function of $\tau$. Dark and light gray shading
          represent the $1\sigma$ and $3\sigma$ confidence intervals,
          respectively.}
	\label{Age_seq}
\end{figure}

We next tested the relationship adopting a variety of maximum
ages. Figure \ref{Age_seq} shows the QR result for clusters with
maximum ages spanning the range from $10^{6.9}$ yr to $10^{7.8}$
yr. In all six test conditions, the values of the maximum cluster
masses appear to be functions of galactocentric radius. For small
$\tau$, the slope fluctuates around 0, while for $\tau \approx 0.8$,
the slopes increase. No matter the extent to which we adjust the age
range, the intercept remains the same and the extent to which the
slope fluctuates remains stable. A zero slope is, in all cases,
contained within the $1\sigma$ range. It is thus clear that varying
the upper age limit has a negligible effect on the cluster
mass--galactocentric radius relation.

In addition, we point out that the uncertainty becomes rather large
for $\tau \geq 0.9$. This strongly suggests that the stochastic ICMF
does not depend significantly on the upper age limit adopted. As long
as the clusters are sufficiently young, stochastic effects appear to
average out.

\subsection{Lower Mass Limit}

\begin{figure}[htbp]
	\centering
	\subfigure[$\log(M_{\rm cl}/M_\odot) \ge 3.3$]{
		\includegraphics[scale=0.3]{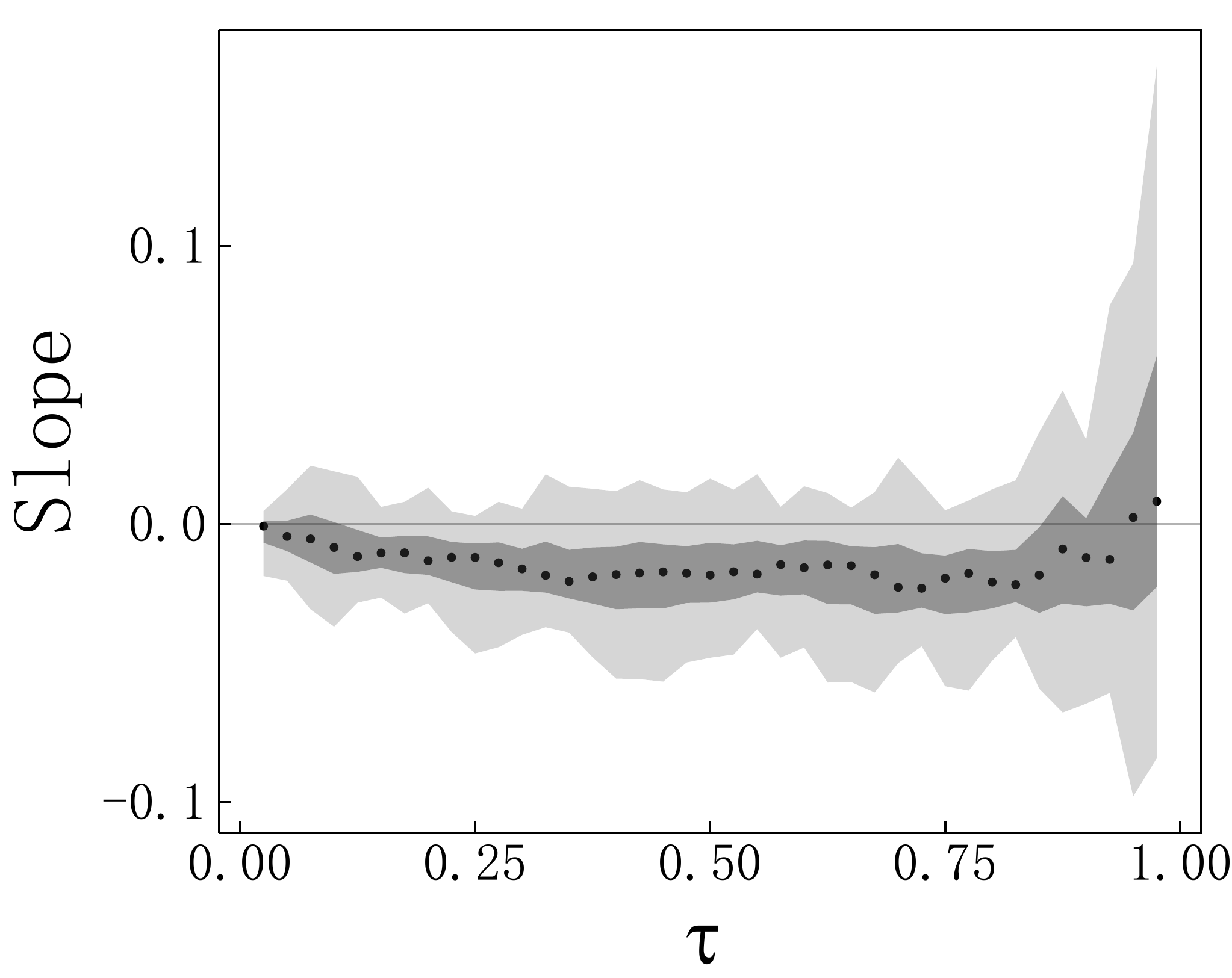} 
	}
	\subfigure[$\log(M_{\rm cl}/M_\odot) \ge 3.5$]{
		\includegraphics[scale=0.3]{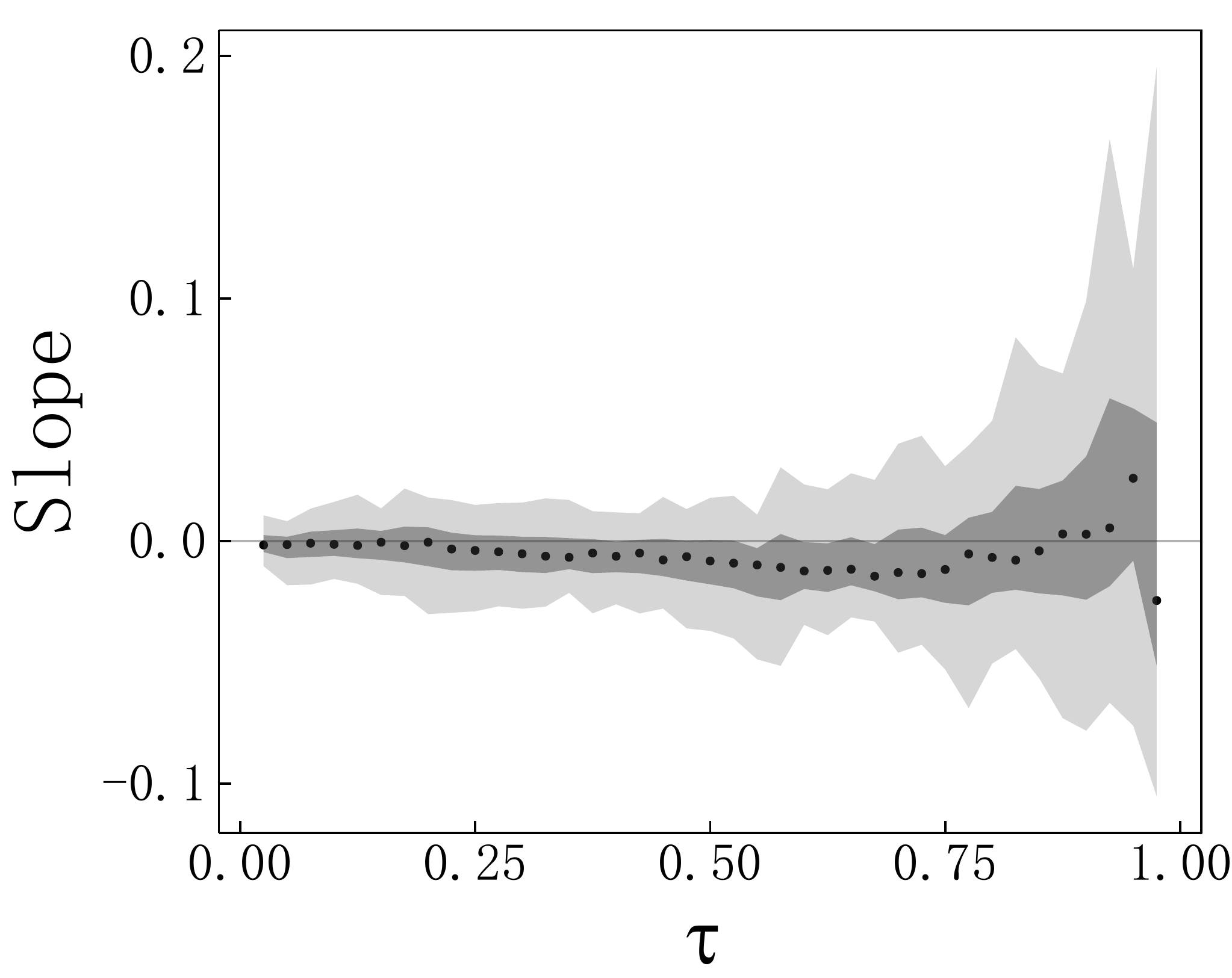} 
	}
    \\
	\subfigure[$\log(M_{\rm cl}/M_\odot) \ge 3.7$]{
		\includegraphics[scale=0.3]{M51_QR_7_3-7.pdf} 
	}
	\subfigure[$\log(M_{\rm cl}/M_\odot) \ge 3.78$]{
		\includegraphics[scale=0.3]{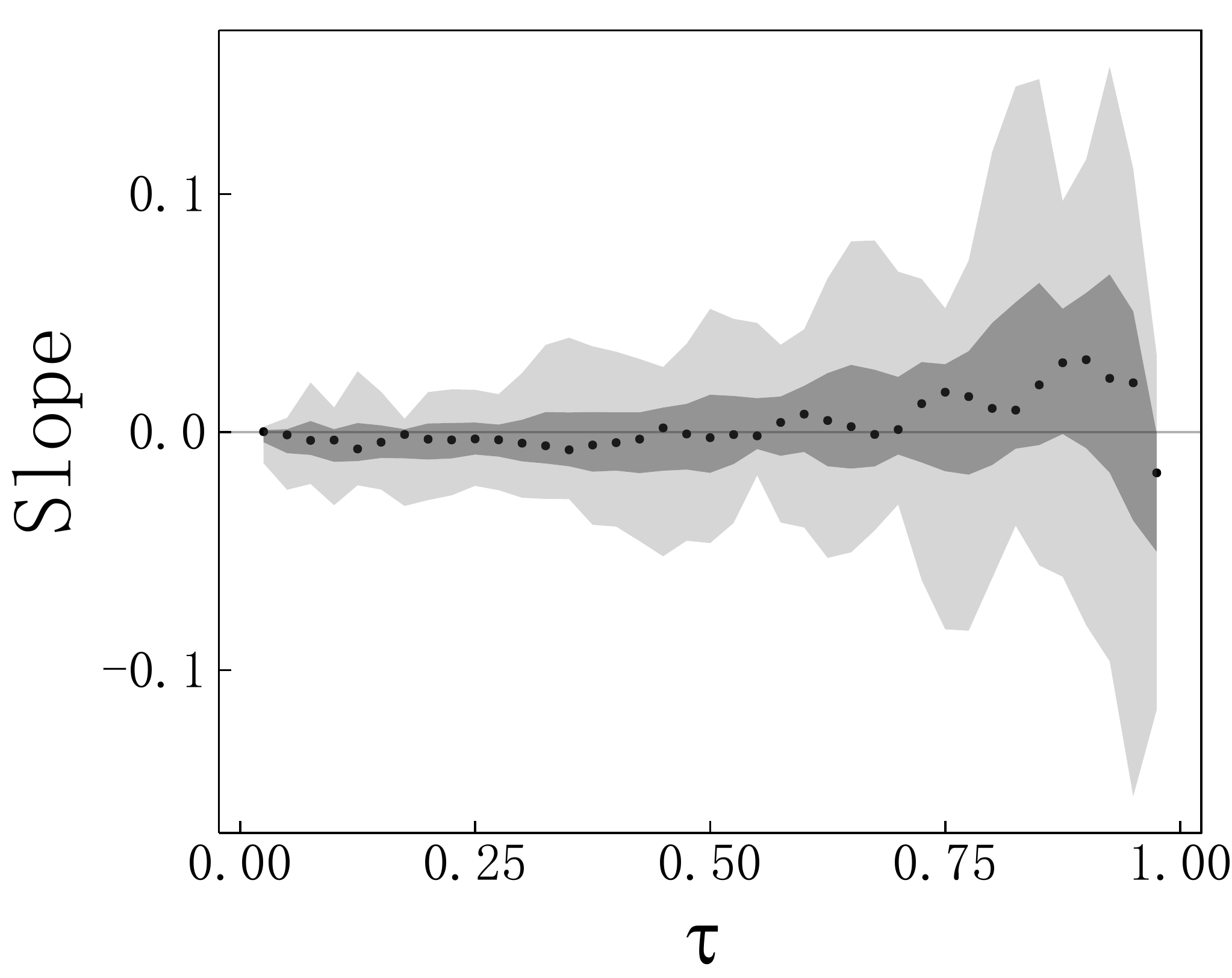} 
	}
	\caption{As Fig. \ref{Age_seq}, but for different minimum mass
          limits, as indicated.}
	\label{Mass_seq}
\end{figure}

Figure \ref{Mass_seq} is similar to Fig. \ref{Age_seq}, but for
variations in the minimum cluster mass limit adopted; we varied the
lower mass limit from $10^{3.3} M_\odot$ to $10^{3.9} M_\odot$ to
assess the effects of sampling incompleteness. The lower mass limit
adopted has a large influence on the results of our fits. In
Fig. \ref{Mass_seq}a, the slope deviates from 0 around the $3\sigma$
confidence level, since the lower-mass limit adopted is well below the
level where our sample is statistically complete. In other cases,
changing the mass limit does not change the resulting pattern
significantly. In Fig. \ref{Mass_seq}b, where we have adopted a
lower-mass limit of $\sim 3200 M_\odot$, the slope is consistent with
a value of 0 at or within the $1\sigma$ confidence level, while
Fig. \ref{Mass_seq}c (for a lower-mass limit of $5000 M_\odot$)
reveals that for all values of $\tau$ the slope is consistent with
zero well inside the $1\sigma$ limit. This leads us to suggest that
the sample is most likely statistically complete for lower masses in
the range from apprimately $3500 M_\odot$ to $5000 M_\odot$. Figure
\ref{Mass_seq}d shows the results for a lower-mass limit of $6000
M_\odot$ as adopted by Chandar et al. (2011; their $\sim 90$\%
completeness limit), which may indeed be a somewhat conservative
choice. Nevertheless, for reasons of consistency with previously
published results, we followed Chandar et al. (2011) as regards the
completeness limit adopted in this paper.

These results can be understood by considering the distribution of the
lowest cluster masses as a function of distance from the center of
M51. For a radial range from the galactic center out to $R \simeq 6$
kpc, and adopting bins of equal radial ranges for simplicity, the {\it
  numbers} of clusters with masses $\log(M_{\rm cl}/M_\odot) \ge 3.7$
do not exhibit any significant radial trend, irrespective of the
radial bin size adopted (for comparison, see Table 1; although there
we included the full range in galactocentric distances covered by our
sample clusters). However, if we adopt a lower-mass limit of
$\log(M_{\rm cl,min}/M_\odot) = 3.3$, an overall downward radial trend
becomes discernible, despite the significant fluctuations on interarm
scales. Our simulations involving mock clusters lead to a similar
result. If the lower-mass limit adopted is sufficiently high enough so
that the level of sampling completeness is $\gtrsim 80$\%, the
artificial clusters can be fitted adequately with a Schechter MF. For
lower completeness fractions, the fits become significantly worse.

\subsection{Radial Bin-size Variations}
\begin{figure}[htbp]
  \centering 
  \subfigure[$N = 40$]{\includegraphics[width=0.38\textwidth]{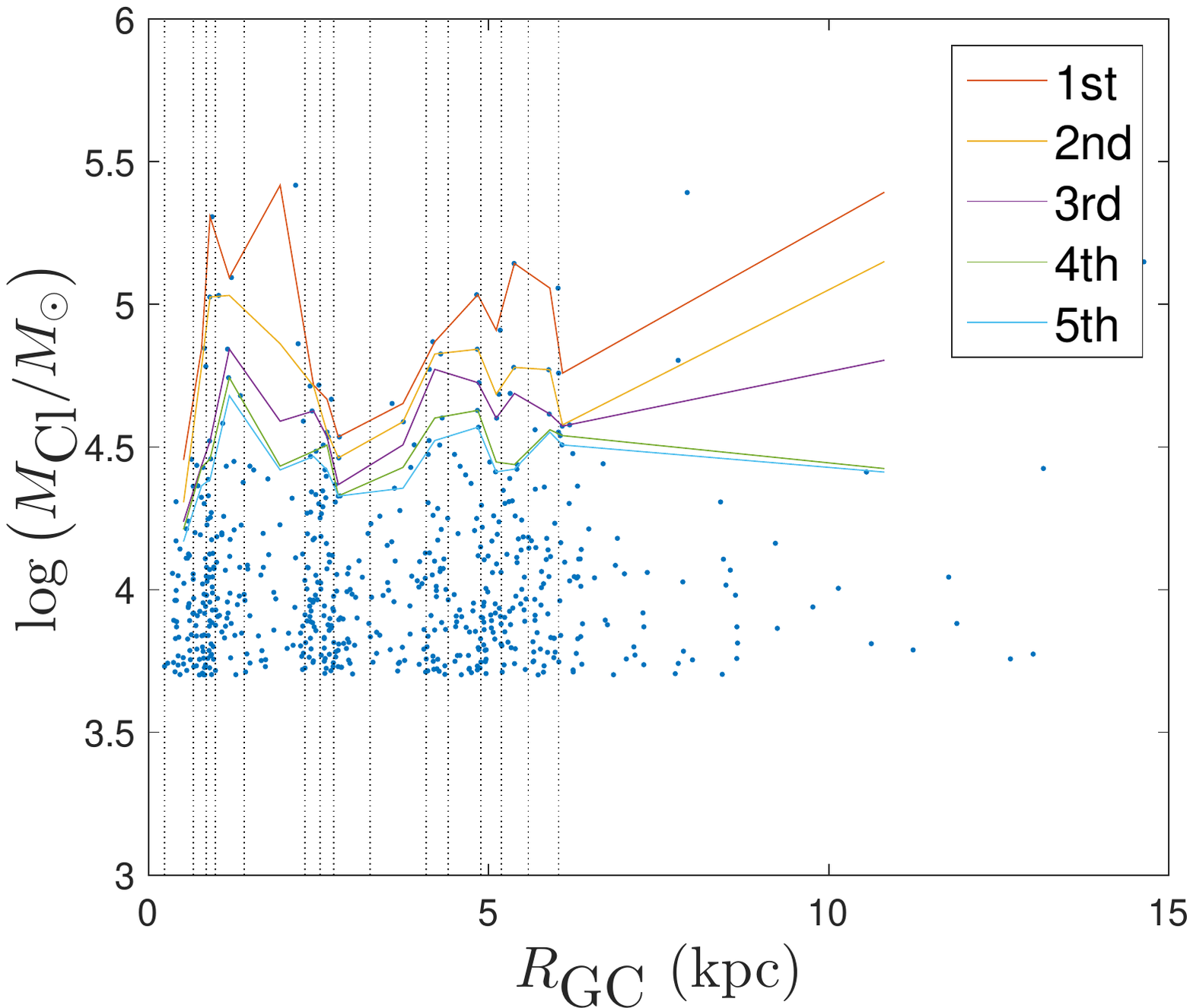}}
  \subfigure[$N = 50$]{\includegraphics[width=0.38\textwidth]{M51_rad-mass_7_3-7_50.pdf}}
  \\ 
  \subfigure[$N = 60$]{\includegraphics[width=0.38\textwidth]{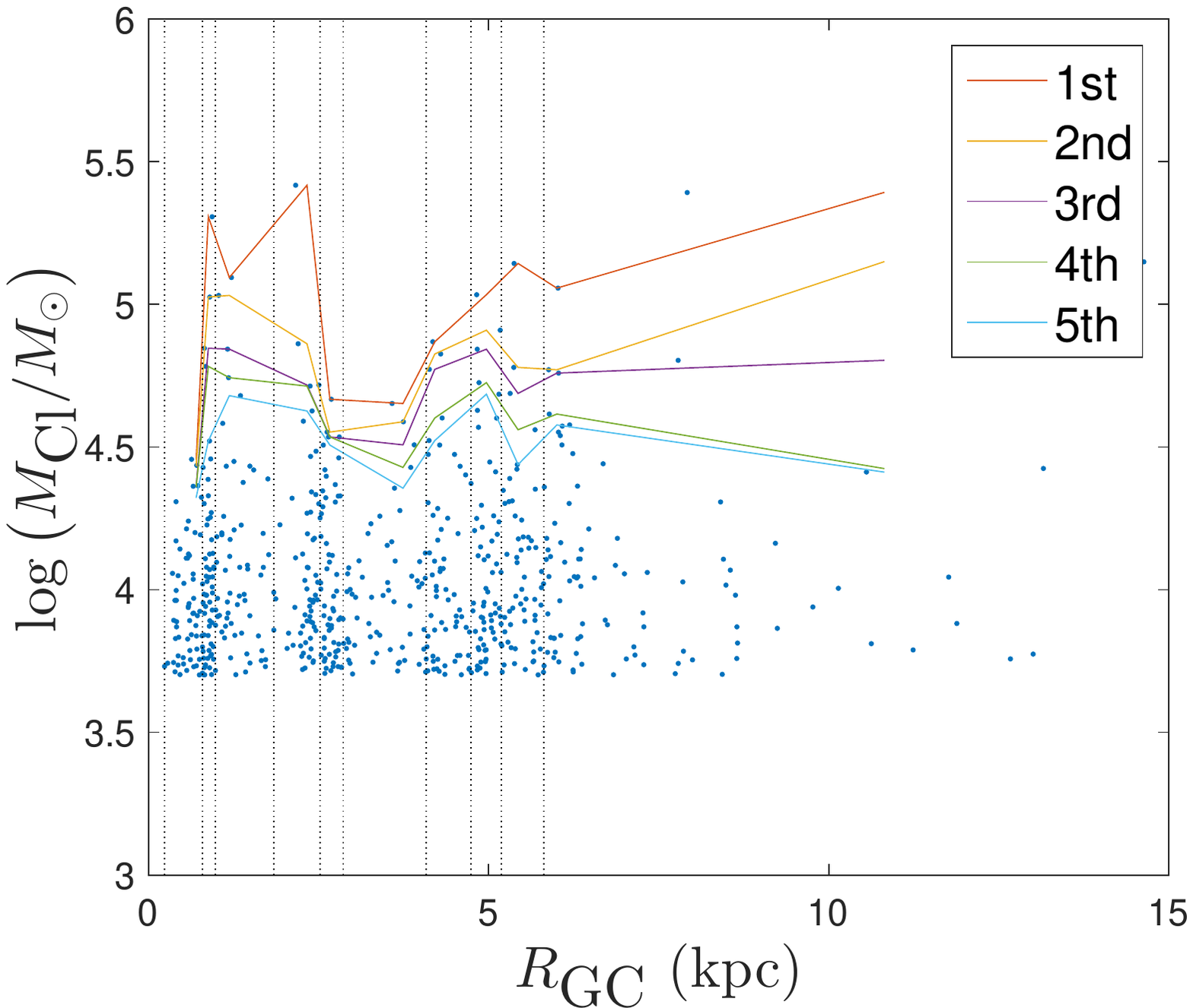}}
  \subfigure[$N = 80$]{\includegraphics[width=0.38\textwidth]{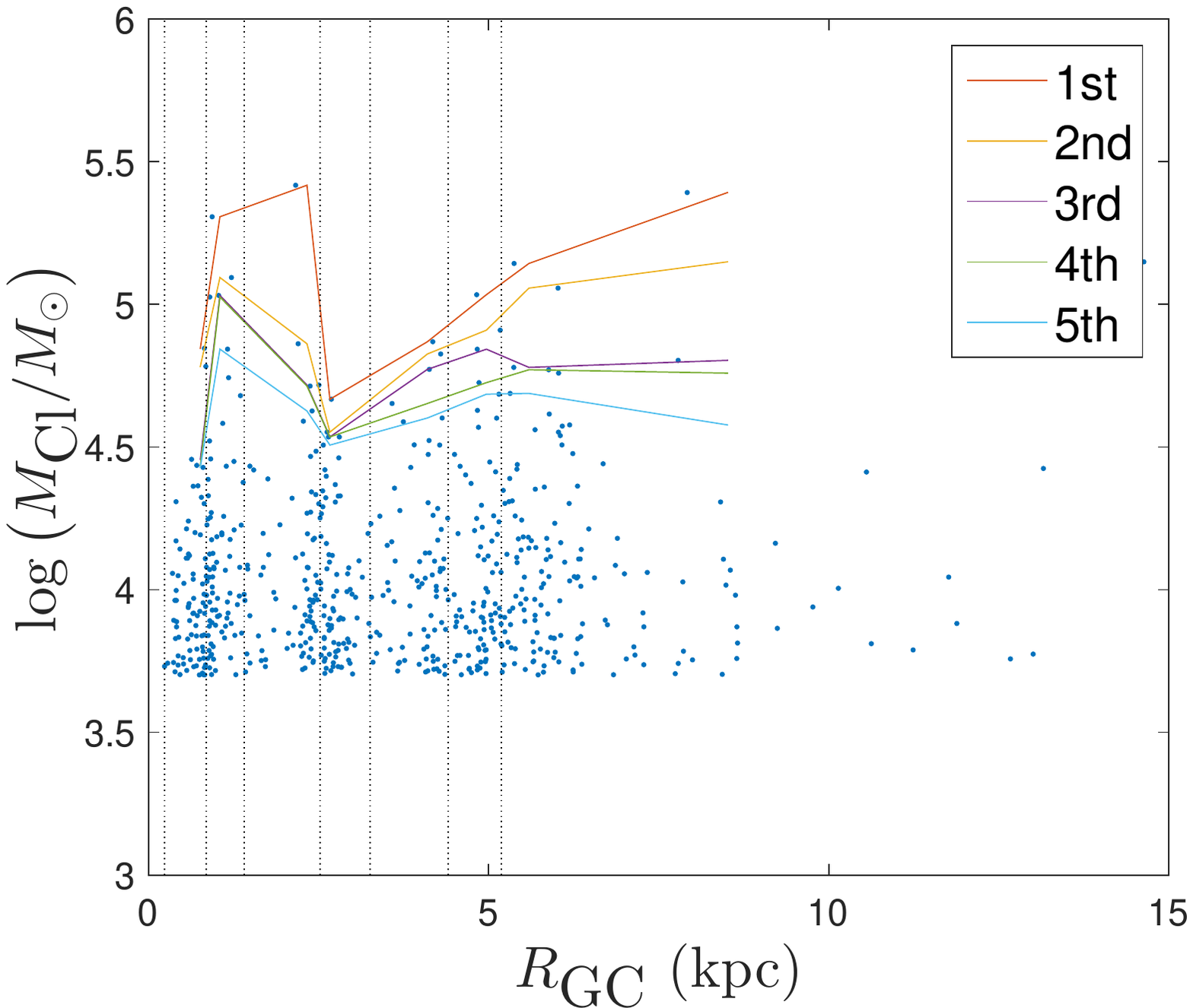}}
\caption{OLS analysis of the M51 star cluster distribution for
  different radial bin sizes, with cluster numbers as indicated.}
\label{Bin_seq}
\end{figure}

In Fig. \ref{Bin_seq}, the number of clusters in each radial bin
varies from 40 to 80. The slopes of the $i^{\rm th}$-ranked lines in
our OLS fits have similar values (see Table \ref{tbl_M51} for
details). Although we see slightly decreasing trends, the values are
consistent with flat slopes, within the uncertainties. This suggests
that the choice of radial bin size does not affect the results
significantly.

\section{The M83 Cluster Population}
\label{M83}

M83 is a barred spiral galaxy, seen on the sky in the constellation
Hydra. We adopted a distance of $D = 4.79$ Mpc
\citep{Karachentsev2007}; the inclination and position angles were
taken as $24^\circ$ and $45^\circ$, respectively \citep{Zimmer2004}.

The majority of the M83 star clusters have ages corresponding to one
of two peaks, at $10^{7.2}$ and $10^{8.1}$ yr. The clusters in M83 are
thus, on the whole, older than those in M51. If we only were to select
clusters associated with the young peak for our analysis, we would be
left with a cluster sample that is insufficiently large to obtain
statistically robust results. We therefore adopted an upper age limit
of $10^8$ yr. The masses of the M83 clusters are also higher than
those in M51. Nevertheless, we still adopt a minimum mass limit of
$M_{\rm cl} = 5000 M_\odot$. Figure \ref{Mass_seq_All}d shows the
radial distribution of the young massive clusters in M83.

\begin{figure}[htbp]
  \centering
	\subfigure[$\log(t \mbox{ yr}^{-1}) \le 7.7$]{
		\includegraphics[scale=0.24]{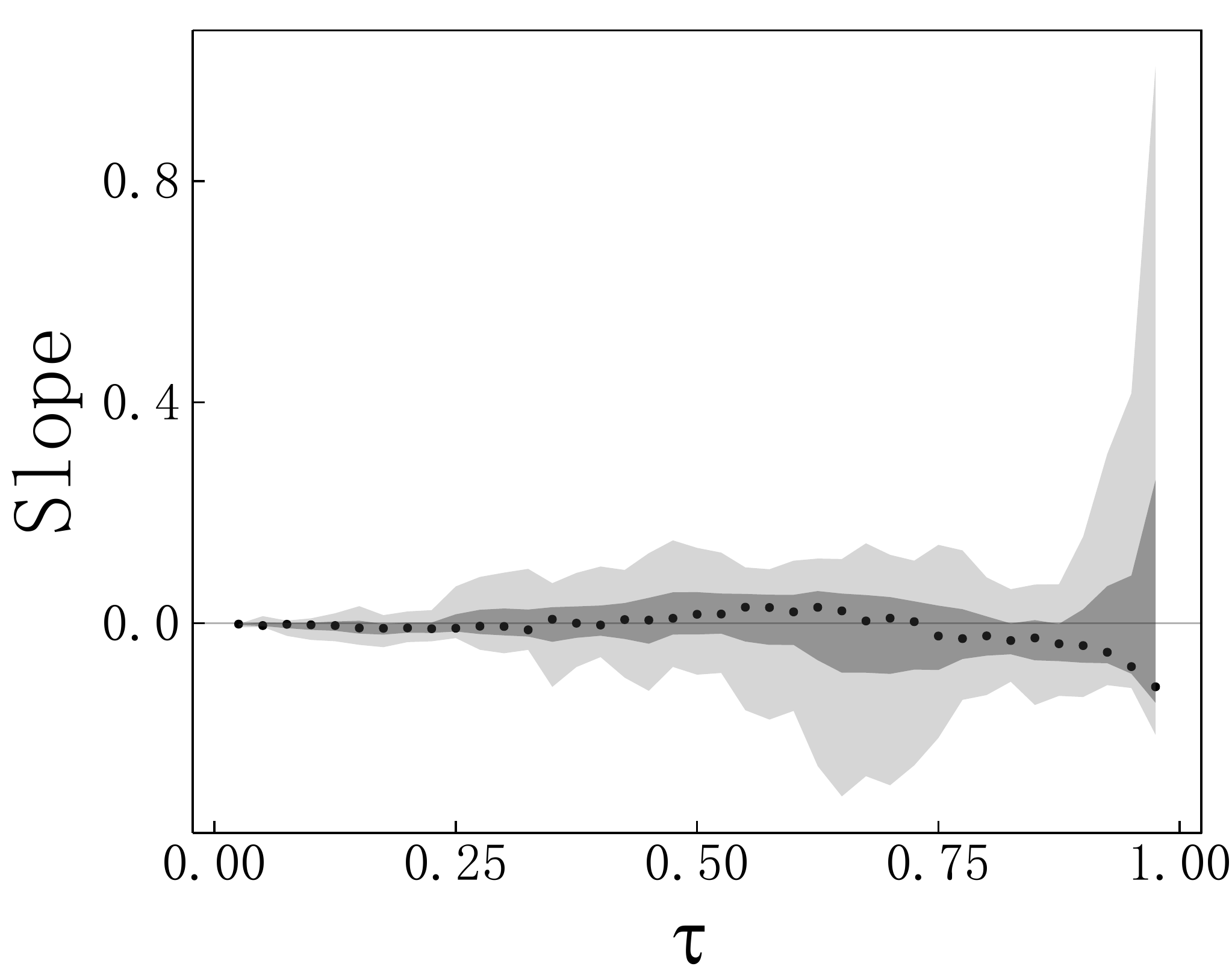}
	}
	\subfigure[$\log(t \mbox{ yr}^{-1}) \le 7.8$]{
		\includegraphics[scale=0.24]{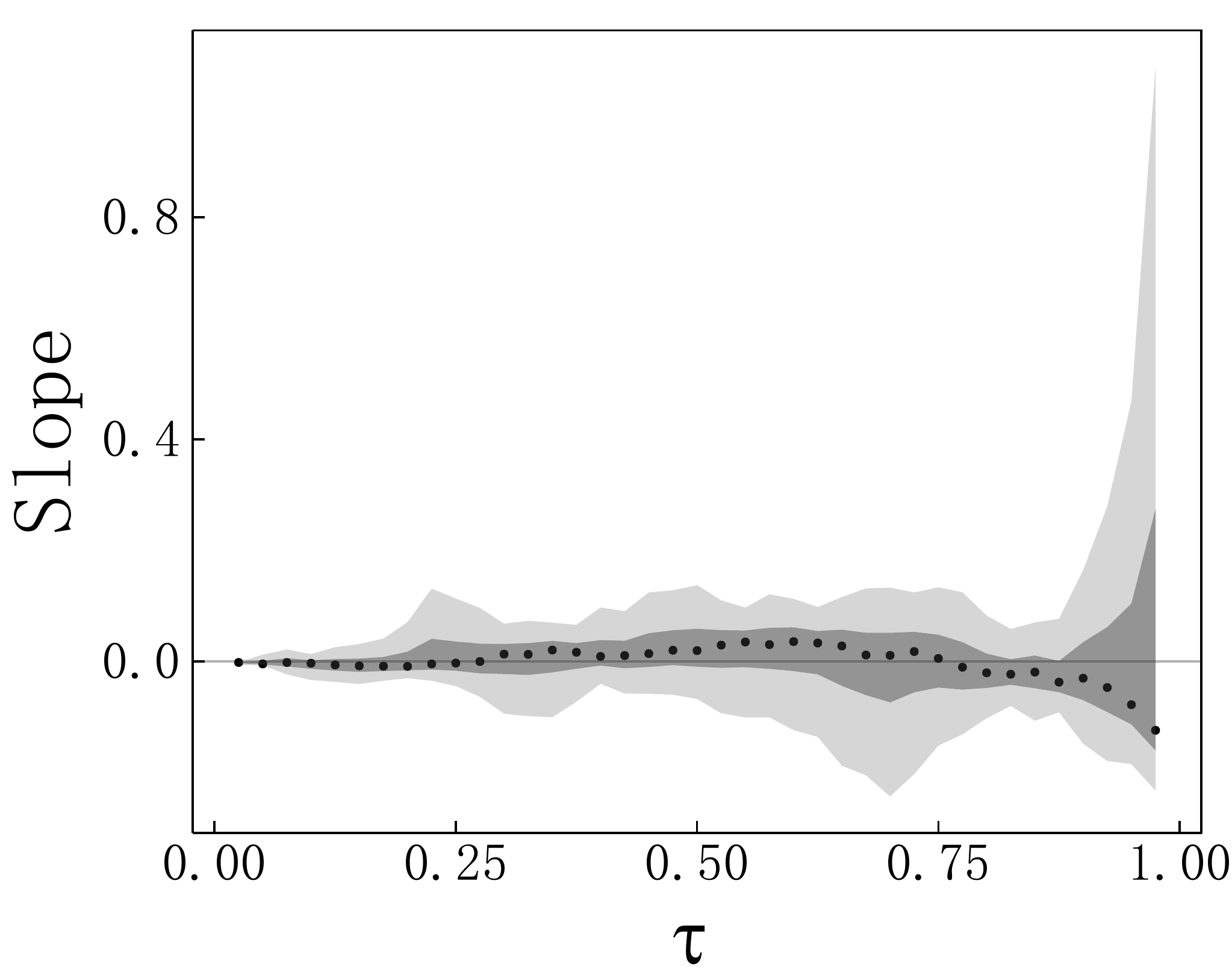}
	}
	\subfigure[$\log(t \mbox{ yr}^{-1}) \le 7.9$]{
		\includegraphics[scale=0.24]{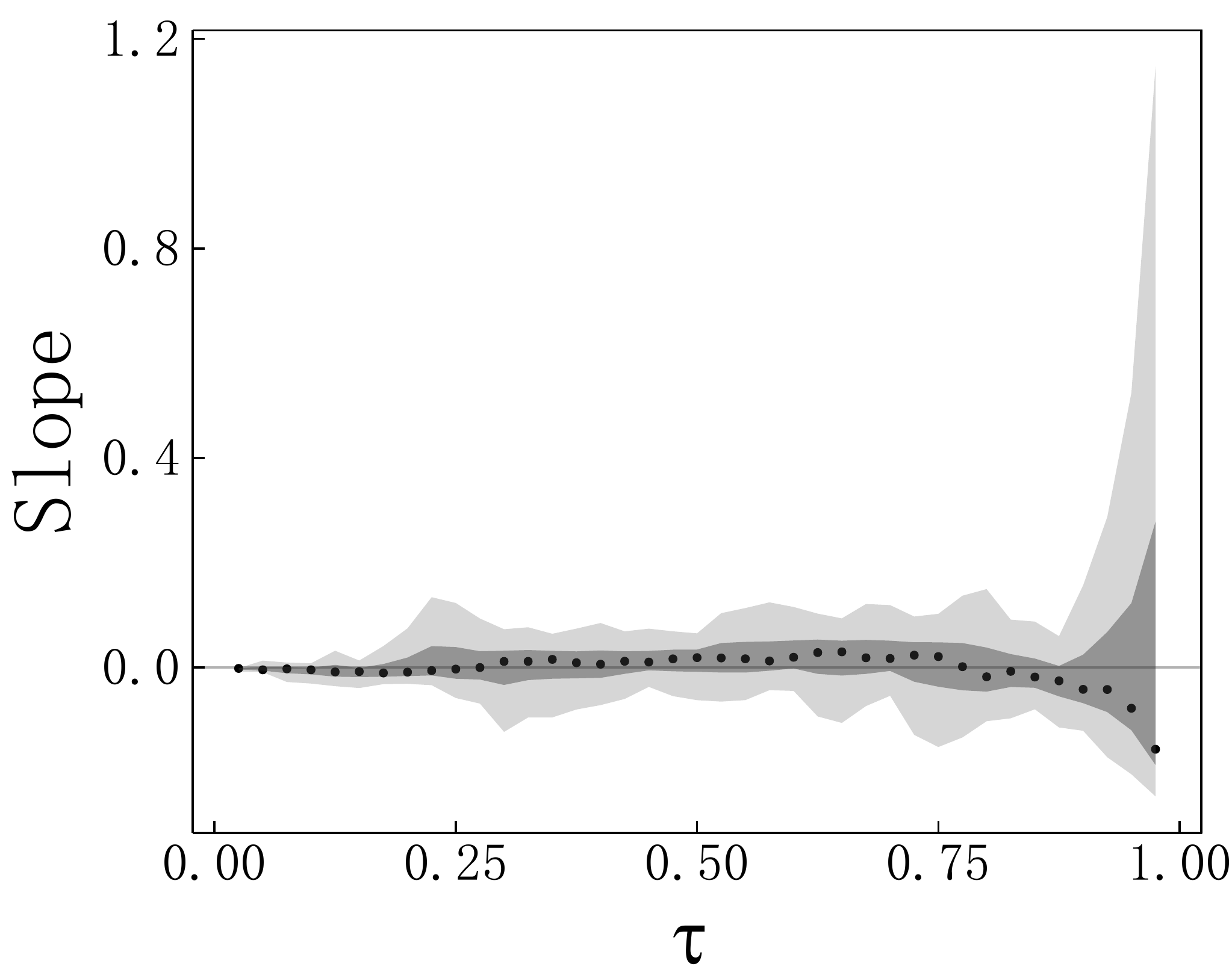}
	}
	\\
	\subfigure[$\log(t \mbox{ yr}^{-1}) \le 8$]{
		\includegraphics[scale=0.24]{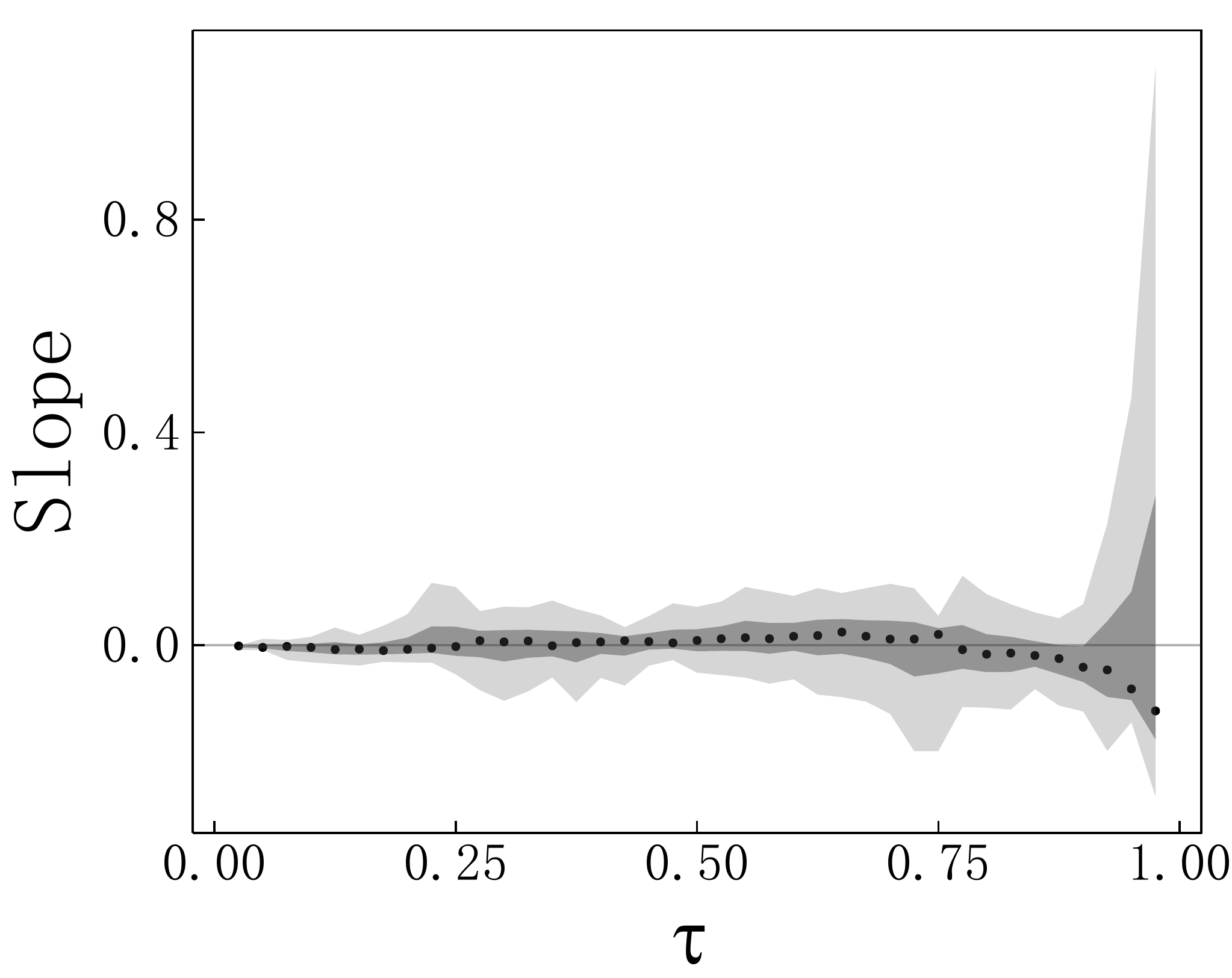}
	}
	\subfigure[$\log(t \mbox{ yr}^{-1}) \le 8.1$]{
		\includegraphics[scale=0.24]{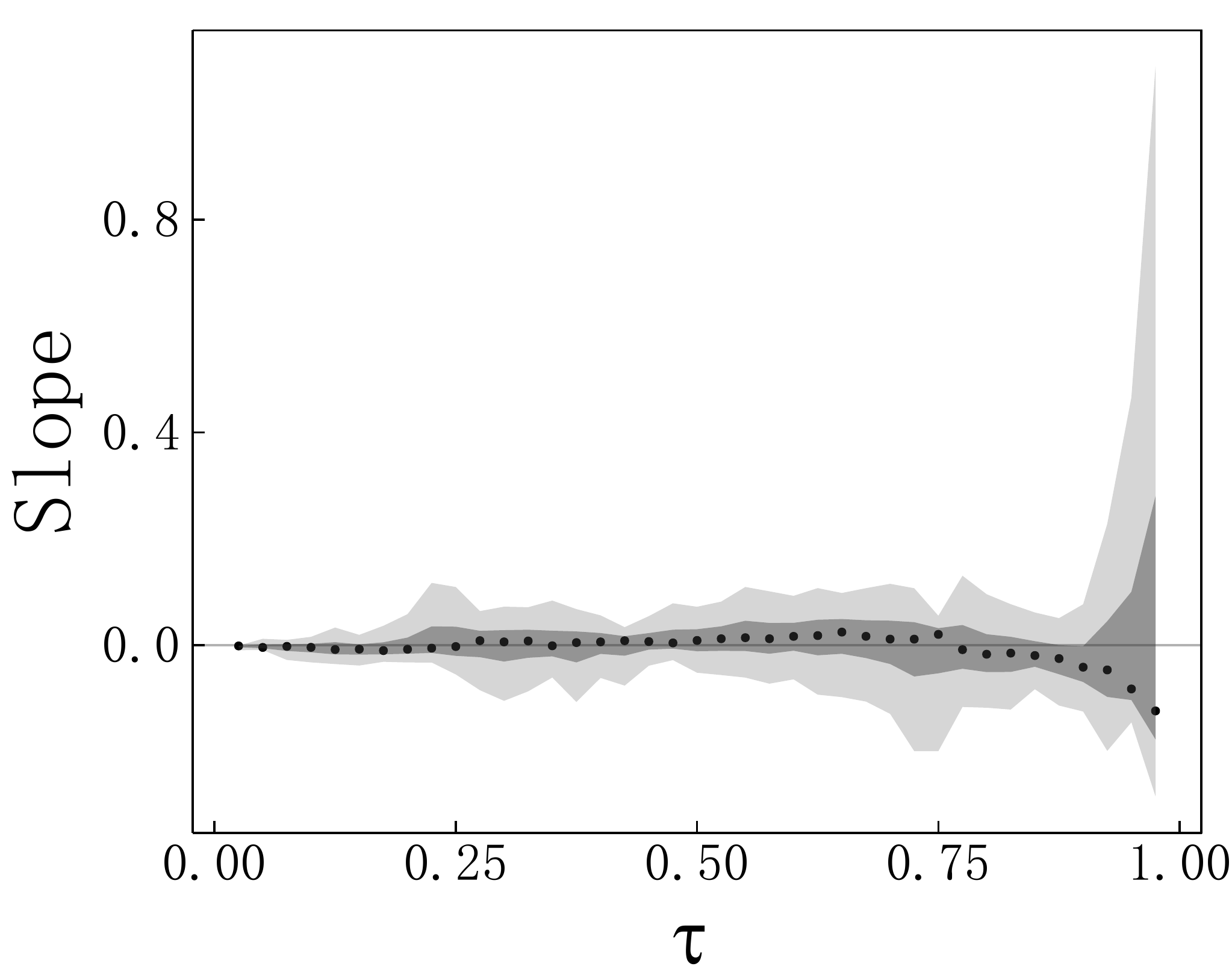}
	}  
	\subfigure[$\log(t \mbox{ yr}^{-1}) \le 8.2$]{
		\includegraphics[scale=0.24]{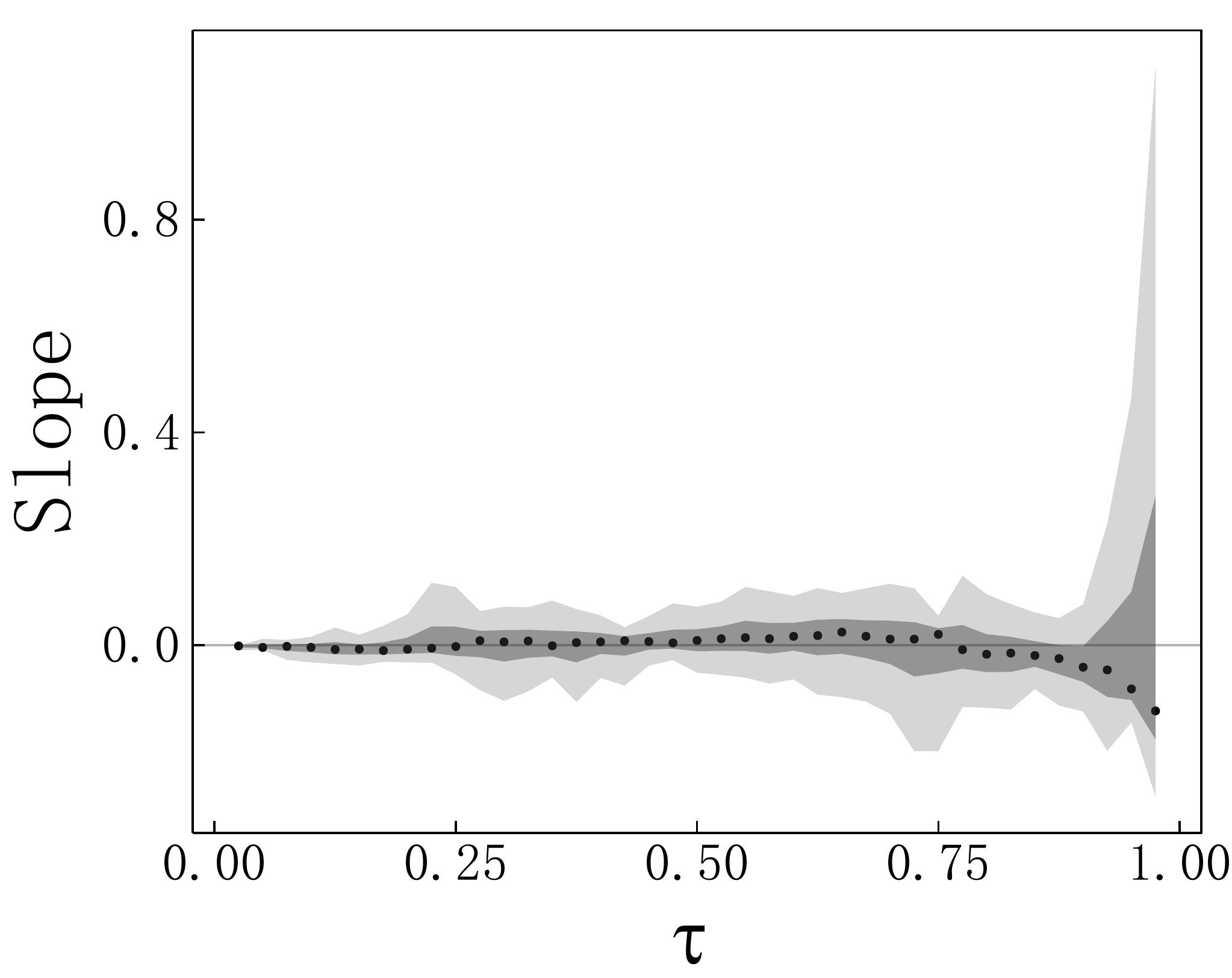}
	}
	\caption{As Fig. \ref{Age_seq}, but for the M83 cluster
  population.}
	\label{Age_seq_M83}
\end{figure}

\begin{figure}[htbp]
  \centering
	\subfigure[$\log(M_{\rm cl}/M_\odot) \ge 3.3$]{
		\includegraphics[scale=0.3]{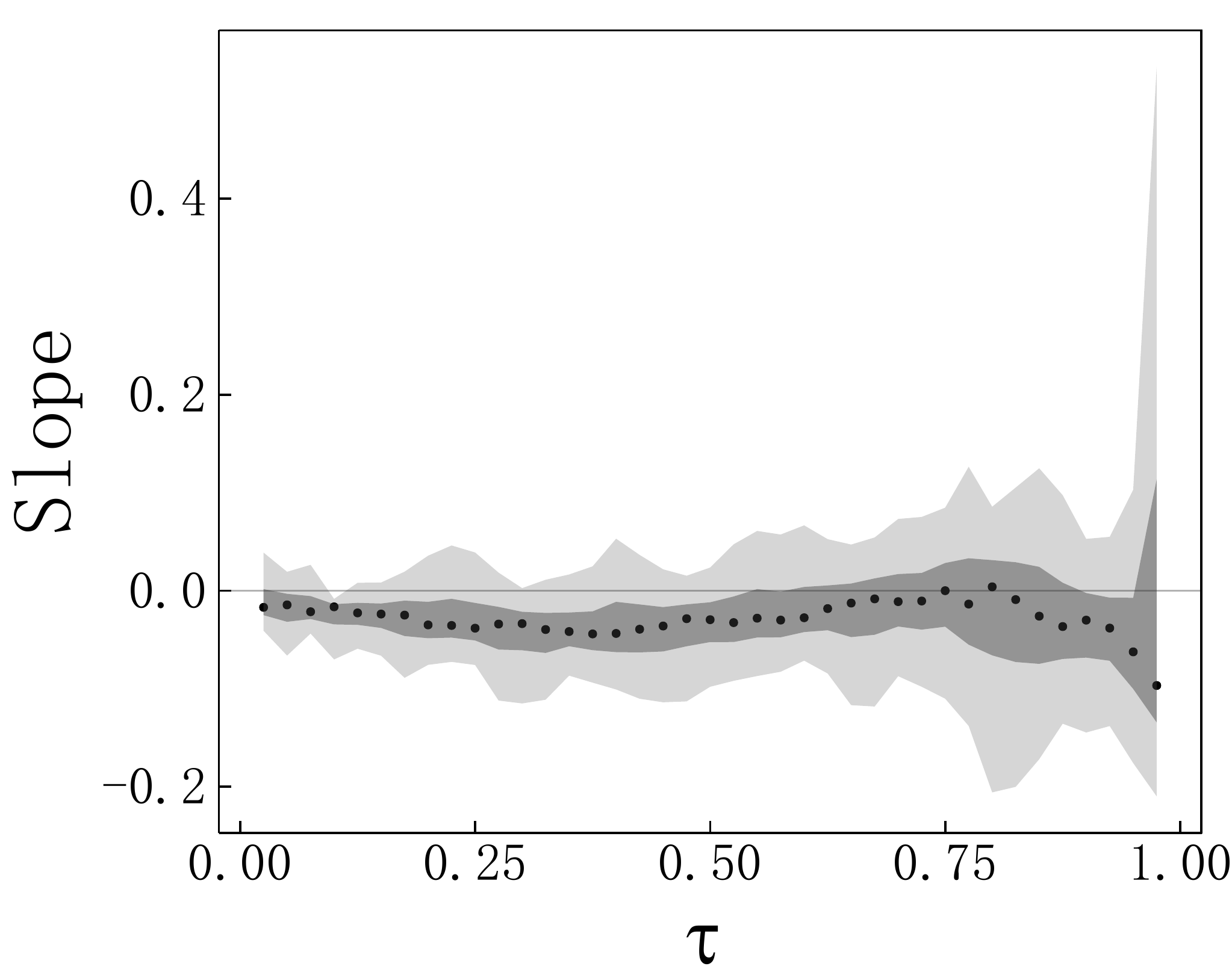} 
	}
	\subfigure[$\log(M_{\rm cl}/M_\odot) \ge 3.5$]{
		\includegraphics[scale=0.3]{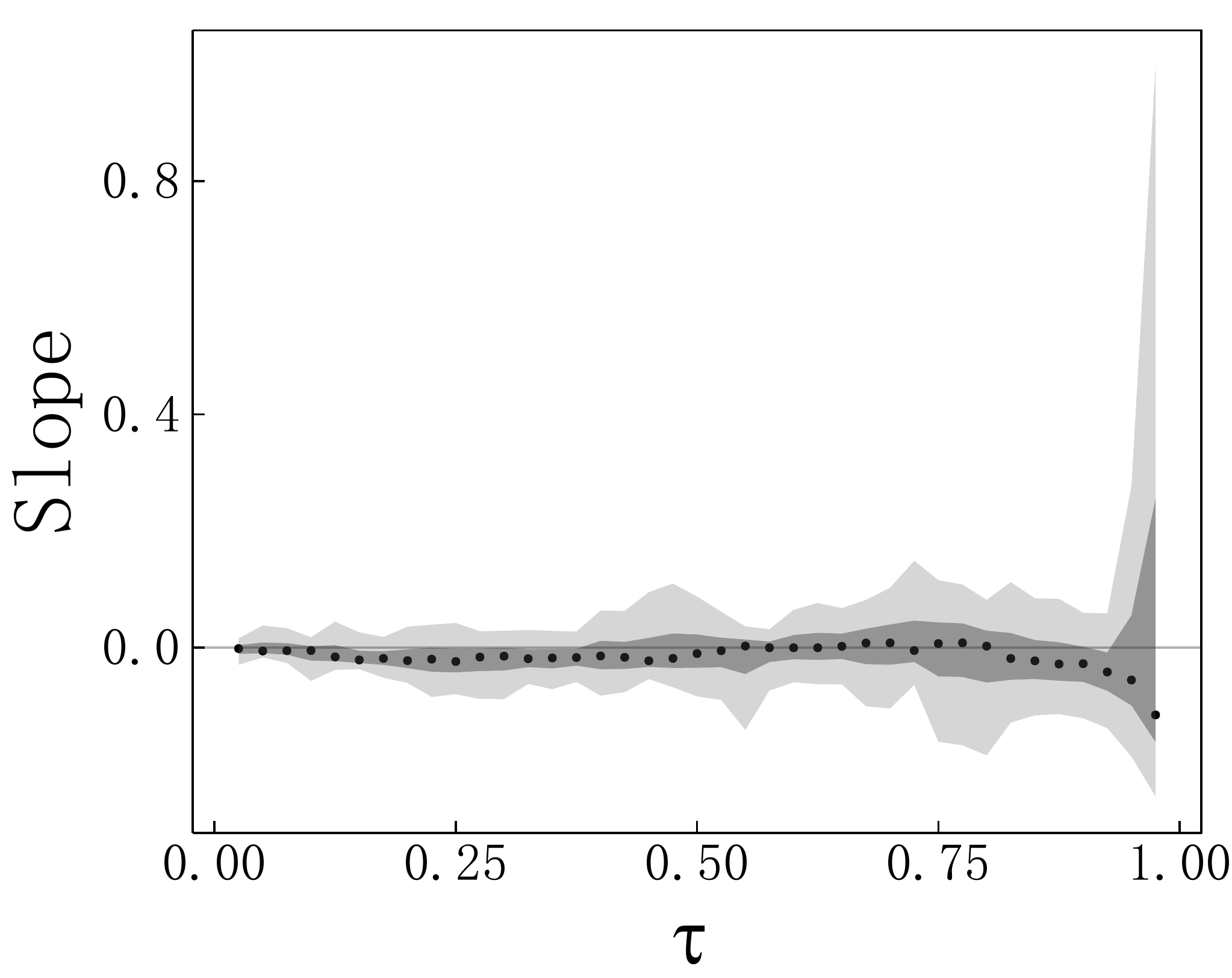} 
	}
	\\
	\subfigure[$\log(M_{\rm cl}/M_\odot) \ge 3.7$]{
		\includegraphics[scale=0.3]{M83_QR_8-0_3-7.pdf} 
	}
	\subfigure[$\log(M_{\rm cl}/M_\odot) \ge 3.9$]{
		\includegraphics[scale=0.3]{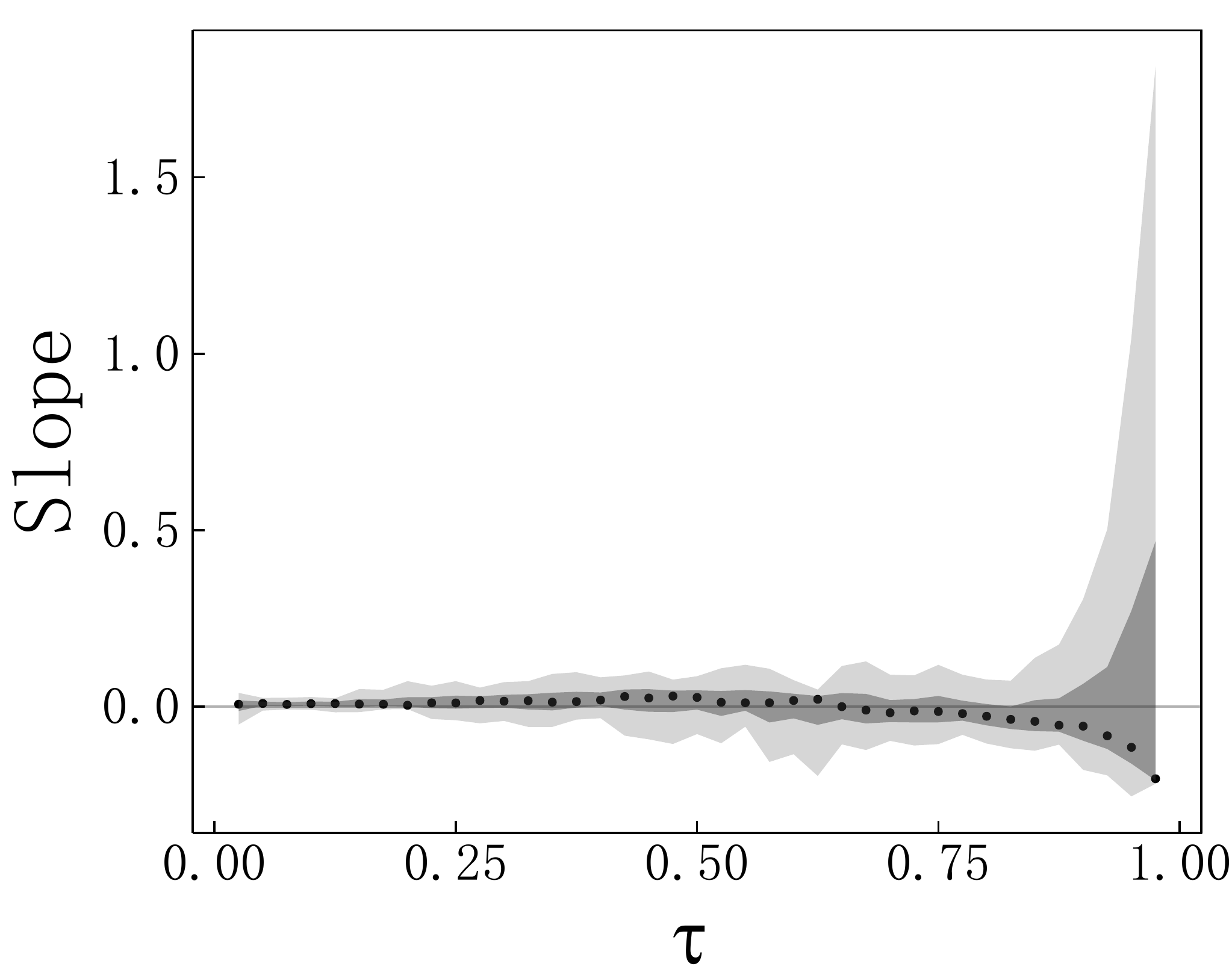} 
	}
	\caption{As Fig. \ref{Mass_seq}, but for the M83 cluster population.}
\label{Mass_seq_M83}
\end{figure}

Figure \ref{Age_seq_M83} is similar to Fig. \ref{Age_seq}. They show
some common ground, although the slope decreases for large values of
$\tau$; the fluctuations of the slope in M83 appear stronger than
those pertaining to the M51 cluster population, particularly for the
larger quantiles. Figure \ref{Mass_seq_M83} is similar to
Fig. \ref{Mass_seq}. Somewhat differently from the results for M51,
the fluctuation level again increases slightly for larger quantiles.

We point out that, as in Fig. 2, for a small number of values of
$\tau$ in both Fig. \ref{Age_seq_M83} and Fig. \ref{Mass_seq_M83}, the
Wald test cannot be used to rule against rejection of the null
hypothesis (zero slope). Again, this can be understood by realizing
that the Wald test is, strictly speaking, only applicable to single
quantiles, while in this case we run into non-independent multiple
hypothesis testing if we are to look over all possible quantiles.

\section{The LMC Cluster Population and a Revisit of the M33 Cluster Sample}
\label{LMC}

\begin{figure}[htbp]
	\centering
	\subfigure[]{
		\includegraphics[scale=0.208]{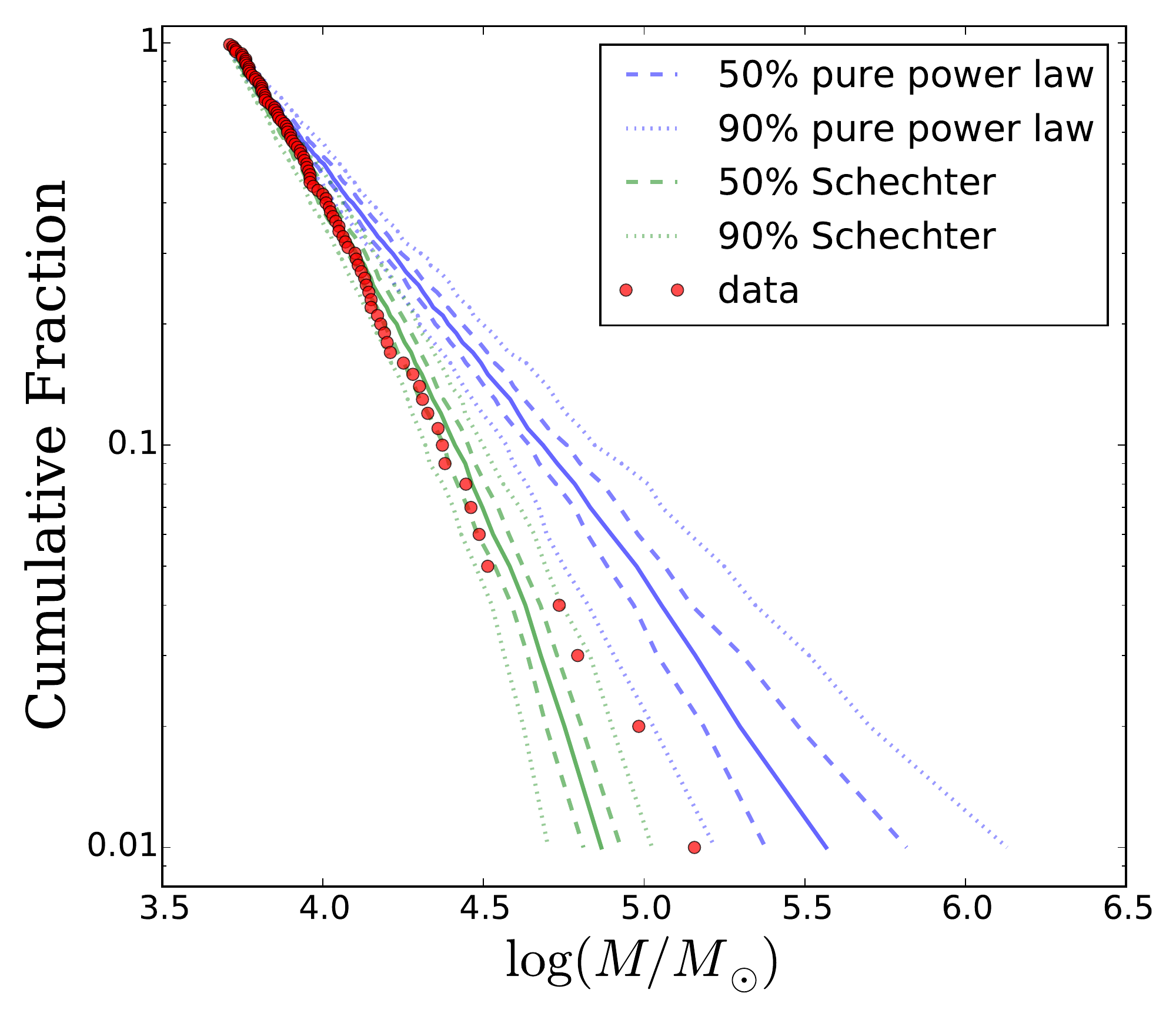} 
	}
	\subfigure[]{
		\includegraphics[scale=0.27]{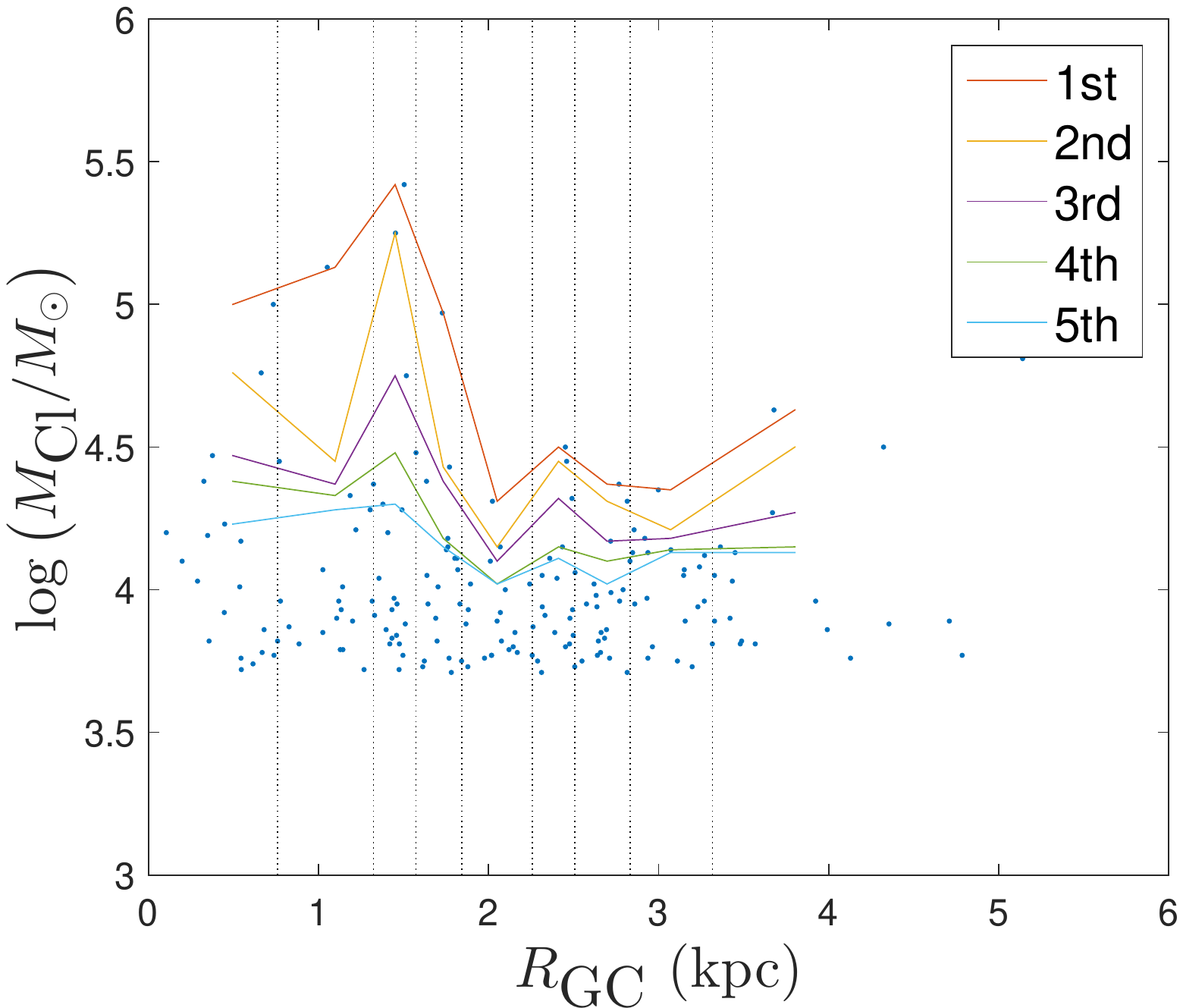} 
	}
	\subfigure[]{
		\includegraphics[scale=0.23]{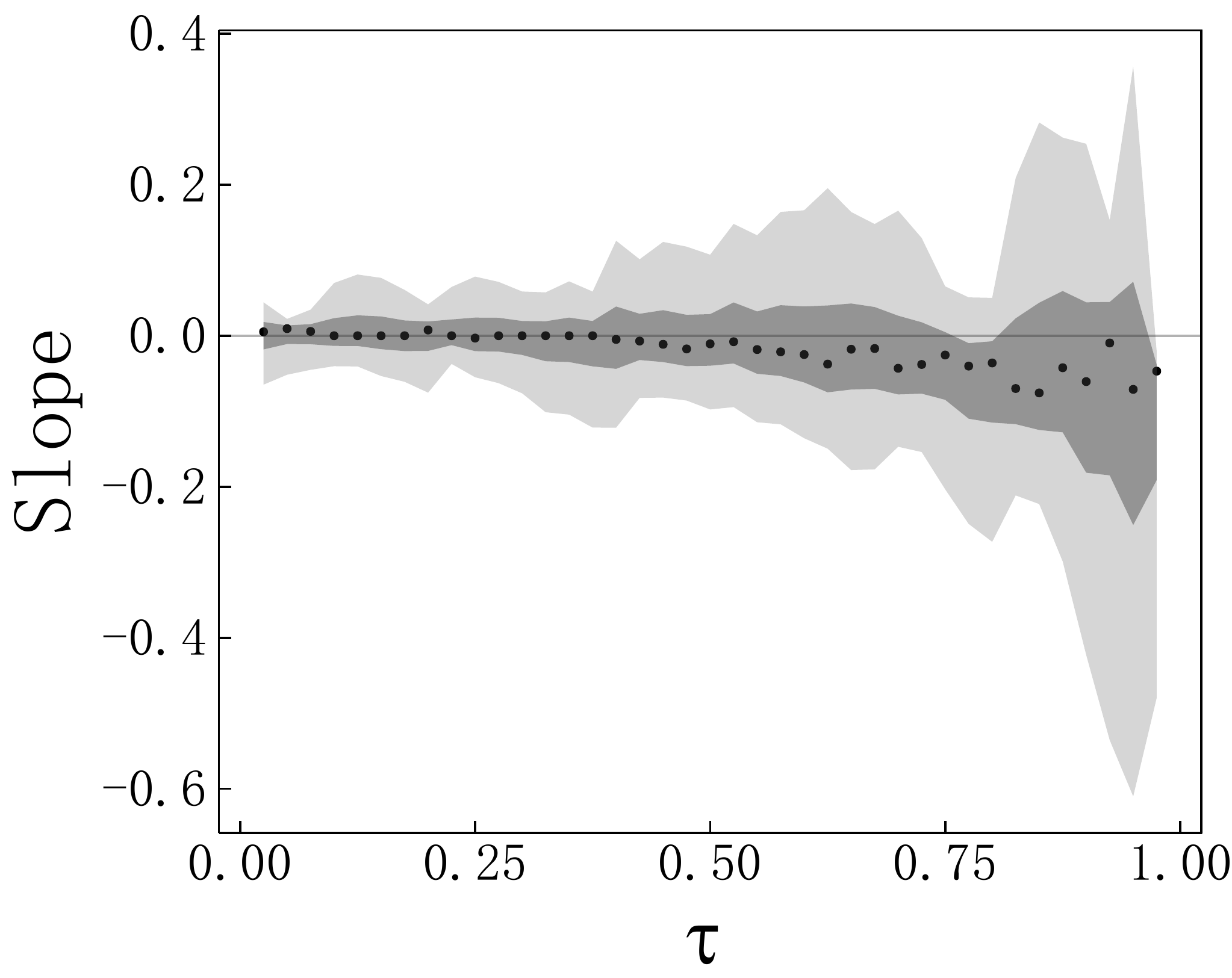} 
	}
        \\
	\subfigure[]{
		\includegraphics[scale=0.208]{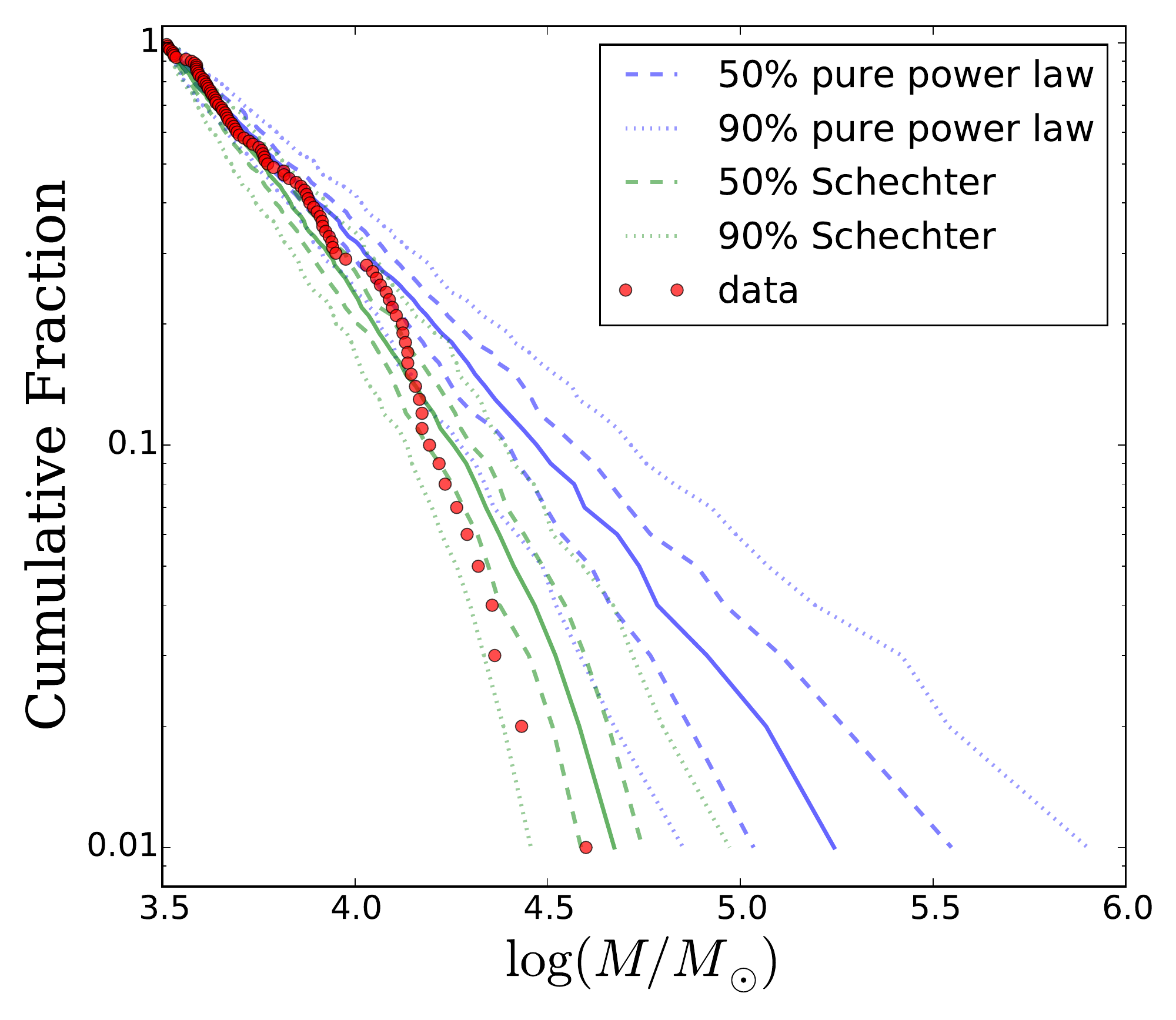} 
	}
	\subfigure[]{
		\includegraphics[scale=0.27]{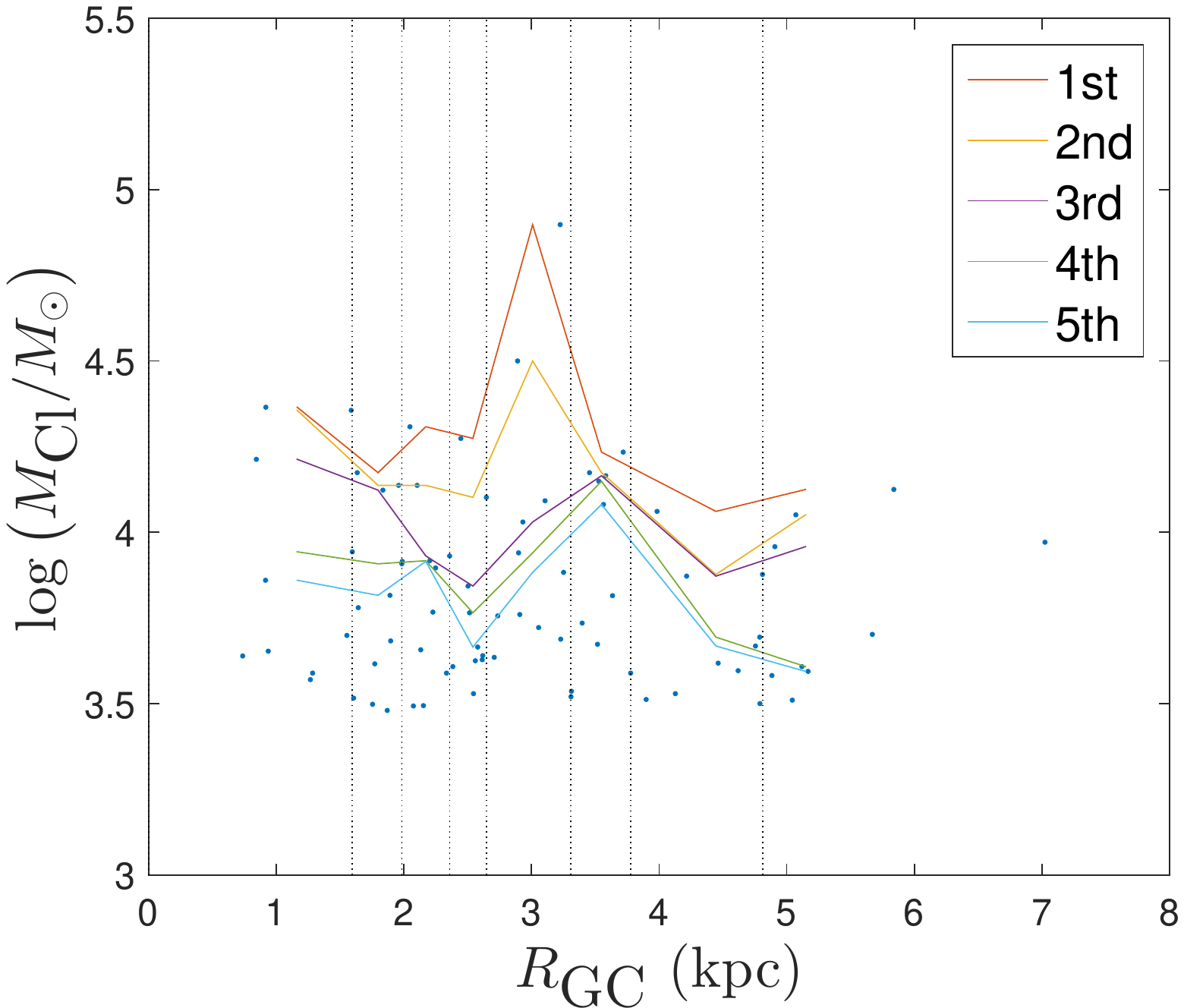} 
	}
	\subfigure[]{
		\includegraphics[scale=0.23]{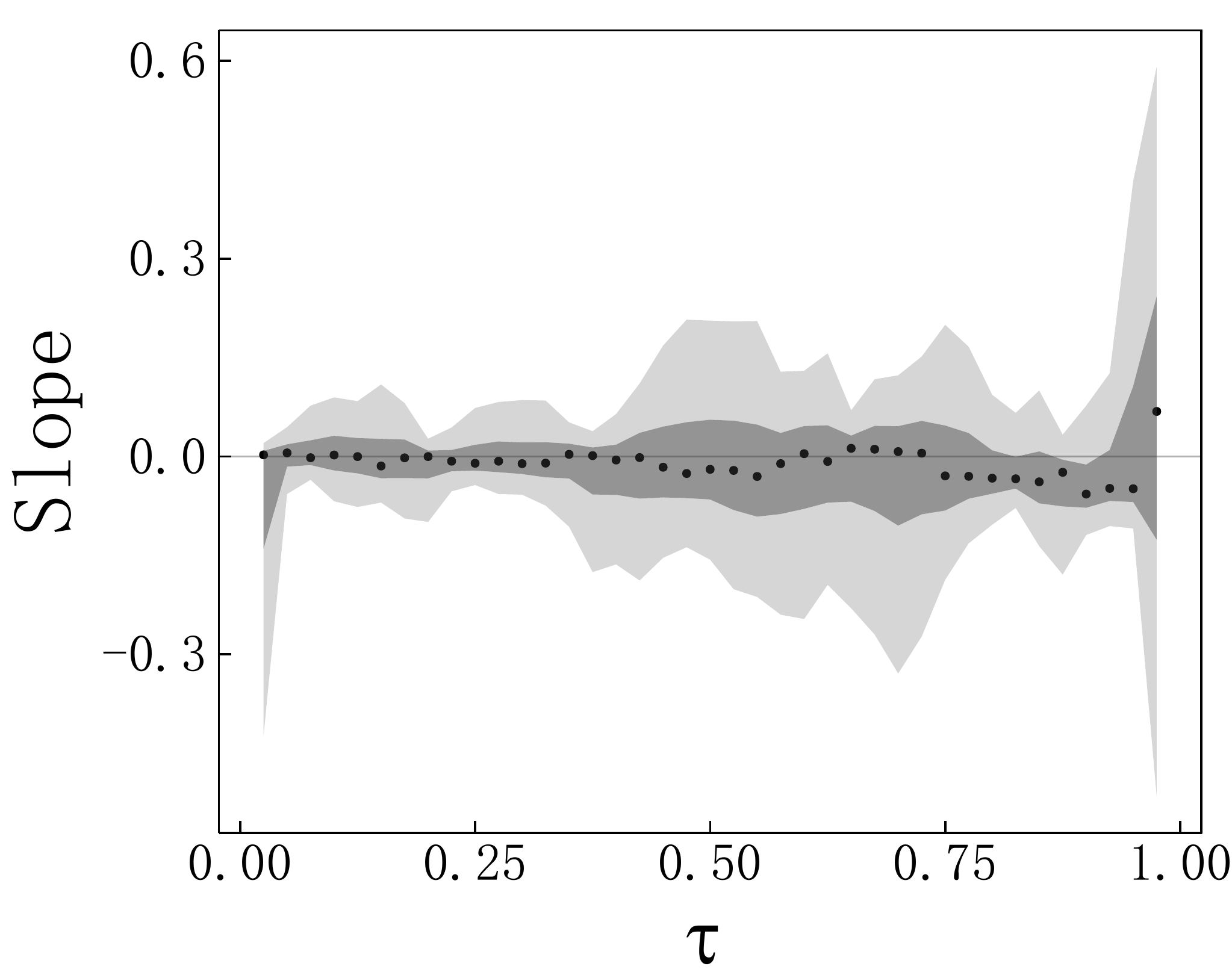} 
	}
	\caption{(a) and (d) Monte Carlo tests applied to (a) the LMC
          and (d) the M33 cluster samples, resulting in a best-fitting
          truncation mass of $6.8 \times 10^4$ $M_\odot$ for a lower
          mass limit of $5000 M_\odot$ in both cases. (b) and (e) OLS
          results based on the cluster mass--galactocentric radius
          distributions of the LMC cluster sample (20 clusters per
          radial bin) and (e) the M33 cluster population (10 clusters
          per radial bin). (c) and (f) QR results for the LMC and M33
          cluster samples, respectively.}
	\label{LMC+M33_all}
\end{figure}

The LMC is an irregular, Magellanic-type dwarf galaxy with a prominent
central bar and hints of spiral arms. We adopted a distance of $D =
49.9$ Mpc \citep{deGrijs2014}. The inclination and position angles
were taken as $34.7^\circ$ and $122.5^\circ$, respectively
\citep{vanderMarel2001}. The majority of the LMC star clusters have
ages corresponding to a dominant peak at $10^{8.0}$ yr. In fact, the
clusters in the LMC have similar age and mass distributions as those
in M83. We adopted an upper age limit of 500 Myr and a minimum mass
limit of $M_{\rm cl} = 5000 M_\odot$ based on Monte Carlo tests (see
Fig. \ref{LMC+M33_all}a). This choice left us with a sample of 179
star clusters. The truncation mass in the LMC is consistent with that
found by Maschberger \& Kroupa (2009).

The linear fit results (see Fig. \ref{LMC+M33_all}b) are included in
Table \ref{tbl_LMC}. Except for the most massive clusters in each
radial bin, the slopes are very small. A more in-depth analysis of the
dataset shows that any relationship between cluster mass and
galactocentric radius is weak. This is exemplified by the QR analysis
shown in Fig. \ref{LMC+M33_all}c.

We next decided to revisit the M33 cluster population. We extended the
age range of interest to cluster ages $\la 500$ Myr so as to check for
environmental effects within M33, if any, on a statically sound
basis. In Fig. \ref{LMC+M33_all}d--f we show the results for this
cluster sample's best-fitting truncation mass and lower mass
limit. Again, although statistically less robust than the results for
either M51 or M83, the behavior of the M33 cluster population
corroborates the scenario deemed most viable for those larger
galaxies.

\begin{table}
\caption{Slopes of the radial distribution of the $i^{\rm th}$ ranked
  most massive star cluster in each bin in the LMC.}
\label{tbl_LMC}
\begin{center}
\begin{tabular}{ccccc}
\tableline\tableline
Condition                                  & $i^{\rm th}$ & Slope    & Intercept & $p$ value \\
                                           &              & ($a$)    & ($b$)     &         \\
\tableline
\multirow{5}{*}{}
$N = 16$                                   & 1            & $-$0.26  & 5.2       & 0.051   \\
$\log\left( M_{\rm{cl}} / M_\odot \right)$ & 2            & $-$0.14  & 4.8       & 0.23    \\
$> 3$;                                     & 3            & $-$0.10  & 4.6       & 0.12    \\
$\log(t \mbox{ yr}^{-1})$                  & 4            & $-$0.095 & 4.4       & 0.054   \\
$\le 8$                                    & 5            & $-$0.058 & 4.3       & 0.092   \\
\tableline
\end{tabular}
\end{center}
\end{table}

\section{Discussion and Conclusions}
\label{Conclusion}

The high-mass regime of the ICMF and its properties have been the
subject of discussions by many authors. Whether or not the ICMF is
established through universal stochastic processes, on scales of
individual galaxies, remains an unsolved problem, however. Young
massive clusters could potentially provide good tools to check the
ICMF's dependence on its environment by means of a careful analysis of
the cluster mass--galactocentric radius relation.

In this paper, we have used star cluster data from M33, M51, M83, and
the LMC to examine the formation scenarios of the young massive star
clusters in these galaxies. By restricting the cluster ages and masses
to within certain limits, we derived the galaxies' cluster MFs. We
also explored the importance of the characteristic ``truncation
mass.'' We first distributed the cluster samples in bins of constant
galactocentric radius. Simplistic application of the OLS method
resulted in a strong trend of decreasing maximum cluster mass with
increasing radius. However, to eliminate size-of-sample effects, we
next adopted bin sizes containing constant numbers of clusters. The
trends pertaining to the first to fifth most massive clusters in each
bin all but disappeared, resulting in near-zero slopes in the context
of our OLS fits.

Second, we applied QR analysis to our data to examine the relation
between the young cluster masses and their galactocentric distances as
a function of the sample quantile, $\tau$. The quantile curves start
from zero for small quantiles ($\tau < 0.2$) and fluctuate for $0.2
\leq \tau \leq 0.8$. For these values of $\tau$, the results are in
agreement with those from the equal-number-binned OLS analysis. Both
methods yield near-zero slopes (within $1\sigma$ to $3\sigma$) in the
cluster mass--galactocentric radius plane. We point out that one
should be careful when directly assessing any environmental dependence
for large $\tau$.

We also investigated the parameter dependence on the maximum age and
minimum mass imposed, as well as that on the bin size adopted. The
resulting slopes do not show any strong dependence on the maximum age
or bin size adopted. However, for M51, the slope decreases
significantly for increasing mass limits. Based on our analysis of the
star cluster populations in M33, M51, M83, and the LMC, we find that
the star cluster mass--galactocentric radius dependence is similar for
all four host galaxies, within the prevailing uncertainties. This
implies that star cluster formation does not necessarily require an
environment-dependent cluster formation scenario, which thus supports
the notion of stochastic star cluster formation as dominant
cluster-formation process (see Gieles et al. 2006; Gieles 2009), at
least within a given galaxy.

\acknowledgements
\label{acknowledgements}

We thank Saurabh Sharma for educating us on the nature of his sample
of M33 objects. We are also grateful to Rupali Chandar for making her
M51 cluster database available, and to the referee for helping us
improve the presentation of our results. W. S. is grateful for support
from Peking University's President's Fund for Undergraduate
Research. R. d. G.  acknowledges research support from the National
Natural Science Foundation of China (NSFC) through grant
11373010. Z. F. is funded through NSFC grant 11373003 and also through
the National Key Basic Research Program of China (973 Program; grant
2015CB857002).

\end{document}